\documentclass[preprint,12pt]{elsarticle}

\usepackage{amssymb}
\usepackage{amsmath}

\usepackage[hyphens]{url}     
\usepackage{breakurl}         
\usepackage{hyperref}         

\usepackage[table,xcdraw]{xcolor}
\usepackage{amsmath,amsfonts,amssymb}
\usepackage{algpseudocode}
\usepackage{algorithm}
\usepackage{booktabs}
\usepackage{multirow}

\journal{Transportation Research Part E: Logistics and Transportation Review}

\begin{document}

\begin{frontmatter}

\title{Collaborative Last-Mile Delivery: A Multi-Platform Vehicle Routing Problem With En-route Charging}

\author[aff1,aff3]{Sumbal Malik\corref{cor1}}
\ead{sumbal.malik@ku.ac.ae}

\author[aff1,aff3]{Majid Khonji\corref{cor1}}
\ead{majid.khonji@ku.ac.ae}

\author[aff1]{Khaled Elbassioni}
\ead{khaled.elbassioni@ku.ac.ae}

\author[aff2,aff3]{Jorge Dias}
\ead{jorge.dias@ku.ac.ae}

\cortext[cor1]{Corresponding author: sumbal.malik@ku.ac.ae, majid.khonji@ku.ac.ae}

\affiliation[aff1]{organization={Department of Computer Science, Khalifa University},
city={Abu Dhabi},
country={UAE}}

\affiliation[aff2]{organization={Department of Computer and Information Engineering}, city={Abu Dhabi}, country={UAE}}

\affiliation[aff3]{organization={KU Center for Autonomous Robotic Systems (KUCARS)},
            city={Abu Dhabi},
            country={UAE}}

\begin{abstract}
The rapid growth of e-commerce and the increasing demand for timely, cost-effective last-mile delivery solutions, have heightened interest in collaborative logistics systems. This research introduces a novel collaborative synchronized multi-platform vehicle routing problem with drones and robots (VRP-DR), where a fleet of $\mathcal{M}$ trucks, $\mathcal{N}$ drones and $\mathcal{K}$ robots, cooperatively delivers parcels. Trucks serve as mobile platforms, enabling the launching, retrieving, and en-route charging of drones and robots, thereby addressing critical limitations such as restricted payload capacities, limited range, and battery constraints. Unlike most existing studies, the VRP-DR incorporates five practical features: (1) multi-visit service, enabling drones and robots to serve multiple customers per trip; (2) multi-trip operations, supporting repeated dispatches; (3) flexible docking, allowing returns to the same or different trucks; (4) cyclic and acyclic operations, enabling return to the same or different nodes; and (5) en-route charging, enabling drones and robots to recharge while being transported on the truck, maximizing operational efficiency by utilizing idle transit time. The VRP-DR is formulated as a mixed-integer linear program (MILP) to minimize both operational costs and makespan. To overcome the computational challenges of solving large-scale instances, a scalable heuristic algorithm, FINDER (Flexible INtegrated Delivery with Energy Recharge), is developed, to provide efficient, near-optimal solutions. Numerical experiments across various instance sizes evaluate the performance of the MILP and heuristic approaches in terms of solution quality and computation time. The results demonstrate significant time savings of the combined delivery mode over the truck-only mode and substantial cost reductions from enabling multi-visits. Additionally, this study provides valuable operational insights into the impact of en-route charging, flexible docking, the number of carrying drones, fleet speed, and payload capacity on the overall performance of the collaborative delivery system.
\end{abstract}

\end{frontmatter}


\section{Introduction}
\label{introduction}

The rapid growth of e-commerce is reshaping urban logistics, bringing significant challenges and opportunities to last-mile delivery systems. According to the World Economic Forum \cite{web1}, the surge in e-commerce is expected to result in a 36\% increase in delivery vehicles within urban areas, with last-mile delivery services projected to grow by 78\% by 2030. As the final step in the retail supply chain, last-mile delivery serves as the primary point of direct customer interaction, but it also poses substantial challenges for logistics providers, often contributing to over 50\% of total logistics costs \cite{kuhn2013integrative}. To remain competitive, companies must meet customer demands for timely and cost-effective deliveries while addressing the growing complexities of urban logistics. Consequently, many enterprises are redesigning their operational models, integrating innovative solutions to improve last-mile efficiency.

Recently, robot-based delivery has emerged as a promising solution, offering a potential transformation of traditional logistics practices. Companies such as Starship Technologies \cite{thevergeThousandsAutonomous} and JD.com \cite{techinasiaTechAsia} have successfully deployed autonomous robots in controlled environments like campuses and residential areas, demonstrating their ability to operate in delivery-restricted zones. However, these robots operate at walking speed to ensure safety, which makes them inefficient for medium to long distance deliveries \cite{yu2022electric}. Similarly, drones have attracted considerable attention, with leading companies such as Amazon \cite{aerospacetestinginternationalAmazonPoised}, DHL \cite{web2}, and UPS launching pilot programs for drone-based parcel delivery \cite{macrina2020drone}. Drones offer clear benefits, including autonomy, speed, and the ability to avoid traffic congestion. Despite their advantages, both drones and robots face limitations, such as restricted payload capacities, short operational ranges, and the need for frequent recharging.

To address these challenges, logistics providers are increasingly exploring collaborative delivery systems that integrate either truck-drone or truck-robot. This approach leverages the strengths of each platform, trucks provide long-distance travel and high load capacity, drones offer speed and flexibility, and robots excel in navigating complex urban environments. In this collaborative approach, trucks act as mobile platforms that support drones or robots, enabling efficient, cooperative operations that enhance overall delivery performance.

The academic community has also shown growing interest in this field, with particular focus on the vehicle routing problem with drones (VRP-D). Wang et al. \cite{wang2017vehicle} introduced a model involving trucks and drones, demonstrating the time-saving potential of cooperative delivery compared to trucks alone. Building on this foundation, researchers have proposed various extensions to the VRP-D, incorporating different fleet configurations, fleet roles, and operational constraints. However, majority of the studies have focused on either truck-drone or truck-robot collaborations, neglecting the potential synergies between these two modes. The integration of a multi-platform system comprising trucks, drones, and robots offers a unique opportunities to address diverse logistical needs while maximizing operational efficiency.

In addition to the limited focus on multi-platform research, existing studies often rely on restrictive assumptions. Most existing studies primarily limit drone/robot to servicing a single customer per trip, known as single-visit operations, whereas multi-visit scenarios, in which a drone/robot serves multiple customers in a single trip, have not been extensively studied \cite{wu2022collaborative}\cite{madani2022hybrid}. Additionally, the effects of drone payload and flight duration/distance on energy consumption have been insufficiently addressed in prior research \cite{wu2022collaborative}. Another critical challenge is the need for recharging drones and robots to enable multiple trips, a factor largely overlooked in existing studies \cite{yu2022electric}. Advancements in charging technologies have introduced diverse charging modes, offering new opportunities for operational efficiency. Previous research has proposed that trucks, whether stationary or in motion, can act as mobile charging stations to support en-route recharging for drones and robots \cite{yu2022electric}\cite{luo2024collaborative}. Flexible docking and acyclic operations, where drones and robots can launch from and return to different trucks and different nodes respectively are also considered rarely \cite{masmoudi2022vehicle}.

This research aims to bridge these gaps by introducing a new variant of the VRP that incorporates flexible and practical operational features. The proposed model introduces:
\begin{itemize}
    \item \textit{Multi-visit services:} allows drones and robots to serve multiple customers in a single trip.
    \item \textit{Multi-trip operations:} enables drones and robots to make multiple trips between launches.
    \item \textit{Flexible docking:} permits drones and robots to launch from and return to same or different trucks, optimizing resource utilization.
    \item \textit{En-route charging:} allows drones and robots to recharge partially or fully while the truck continues along its route.
    \item \textit{Fleet synchronization:} ensures coordinated operations, with trucks arriving at designated locations before drones and robots for docking and recharging.
\end{itemize}
The integration of the aforementioned features makes the problem more practical but also significantly more complex. This research introduces a novel collaborative multi-platform Vehicle Routing Problem with Drones and Robots (VRP-DR). The model involves a fleet of $\mathcal{M}$ trucks, carrying $\mathcal{N}$ drones and $\mathcal{K}$ robots, working collaboratively to deliver parcels across a wide region. Trucks serve as mobile platform for launching, retrieving, and recharging drones and robots. Each customer is served exactly once and can be served by either a truck, drone, or robot. The problem is formulated as a mixed-integer linear program with dual objectives of minimizing operational costs and makespan. To address scalability challenges, a simple yet effective heuristic algorithm, FINDER (Flexible INtegrated Delivery with Energy Recharge), is developed to efficiently solve large-scale instances.

This research makes the following key contributions to the literature:

\begin{itemize}
    \item To the best of our knowledge, this is the first study to propose a collaborative multi-platform last-mile delivery system that integrates three platforms: traditional trucks, drones, and robots. The novel VRP-DR framework incorporates advanced features, including multi-trip operations, multi-visit capabilities, the transportation of multiple parcels in a single trip, and en-route charging for drones and robots. These features significantly enhance the system’s practicality while introducing additional complexity, distinguishing this study from existing work.
    \item The VRP-DR is formulated as a Mixed-Integer Linear Programming (MILP) model that captures the collaboration between the three platforms. The model aims to minimize both the makespan and overall delivery costs, providing an exact solution for the proposed problem. The formulation can be solved using any solver like Gurobi for small-scale instances.
    \item A scalable heuristic algorithm, FINDER, is developed to efficiently solve large-scale instances of the VRP-DR within a reasonable timeframe by decomposing the problem into sequential vehicle assignments while incorporating en-route charging capabilities for drones and robots. The heuristic's performance is validated against the exact MILP solution.
    \item Extensive experiments are conducted across different scale of instances to evaluate the solution quality and the computation time of the exact solution and the heuristic algorithm, providing insights into their practical applicability.
    \item Operational insights are derived through experiments assessing the impact of the model’s key features. Additionally, a detailed sensitivity analysis investigates the influence of critical parameters on system performance.
\end{itemize}

The rest of the paper is structured as follows: Section \ref{related_work} provides a review of the relevant literature. Section \ref{methodology} outlines the problem description, underlying assumptions, and the MILP formulation. Section \ref{solution_method} introduces a scalable heuristic algorithm designed to solve large-scale instances of the proposed problem. Section \ref{experimental_results} details the experimental setup, provides a comprehensive analysis of the results, and discusses sensitivity analyses. Finally, Section \ref{conclusion} summarizes the key findings and outlines the future research directions.

\section{Related Work}
\label{related_work}
Collaborative truck-drone delivery systems (CTDDS) have garnered significant research attention for their potential to enhance efficiency in last-mile delivery. This section reviews the relevant literature on CTDDS and highlights the novelty of this study.

The research on the truck–drone routing problem originates from a variant of traveling salesman problem known as the Flying Sidekick Traveling Salesman Problem (FSTSP), first introduced by Murray et al.\cite{murray2015flying}. This problem addresses a scenario where a drone collaborates with a conventional truck to deliver parcels, allowing a launching/retrieving node to be visited only once. A related problem, the Traveling Salesman Problem with Drones (TSP-D), is proposed by Agatz et al.\cite{agatz2018optimization}. It considers the assignment of a single drone to work in collaboration with a single truck to deliver parcels, allowing cyclic drone operation (where the drone’s launching and retrieving nodes may be the same). Another variant, the VRP-D, first proposed by Wang et al.\cite{wang2017vehicle}, explores the coordination of multiple trucks and drones to serve customers. Majority of the studies on FSTSP, TSP-D, and VRP-D aim to minimize delivery completion times, while others prioritize reducing operational costs. Building on these variants, research efforts focus on incorporating additional model features and developing both exact and heuristic solution approaches. Generally, the literature on CTDDS can be categorized into two main categories: (I) the logistic system where drones or robots are solely responsible for delivering parcels, while trucks handle auxiliary tasks such as launching, retrieving, battery swapping and recharging, and (II) the collaborative systems involving both a truck and robot/drone, where both vehicles serve customers with the drone/robot being launched and retrieved from customer nodes visited by the truck. 

Meng et al. \cite{meng2024multi} modelled a single truck-drone routing problem with time windows in the context of last-mile logistics. A truck equipped with a drone is tasked with both pickup and delivery services for a set of customers, aiming to minimize overall costs. The drone operates in an acyclic mode, allowing it to launch and return at different locations, and it is capable of serving multiple customers in a single trip. The problem is formulated as a large-scale mixed-integer bilinear program, with bilinear terms capturing the load-dependent energy consumption of drones. To address computational challenges associated with large-scale instances, a customized adaptive large neighborhood search (ALNS) algorithm is developed. Similarly, Madani et al. \cite{madani2024optimization} studied a single truck-drone routing problem in which the drone can make multiple deliveries during a single trip. Their objective is to minimize total travel costs for both the truck and the drone. The study proposed an integer linear programming (ILP) model for small-scale instances and a variable neighborhood search (VNS) algorithm for larger instances. However, their approach does not incorporate synchronization between the truck and drone or an energy consumption model for the drone, presenting notable limitation for real-world deployment. Yu et al. \cite{yu2022electric} proposed a two-echelon electric van-based robot delivery system that incorporates en-route charging capabilities. In their model, each van carries a single robot, which can access areas restricted to the vans. While in transit, the robots' batteries are recharged, enhancing the overall efficiency of the delivery system. The problem is formulated as a mixed-integer program with the objective of minimizing total travel costs. To solve larger problem instances, an ALNS algorithm is developed. Sensitivity analyses show that en-route charging, combined with increased battery capacities and faster charging rates, can significantly reduce delivery costs. However, the study has several limitations, including its focus on a single van-robot setup, the lack of flexible docking (requiring robots to be retrieved by the same van that deployed them), and experiments that primarily focus on charging aspects.

Ahn et al. \cite{ahn2024operational} developed an optimization model integrating electric trucks and autonomous delivery robots, addressing realistic customer demands estimated using GIS data. The model optimizes vehicle routing and robot allocation with an emphasis on energy efficiency, including battery aging costs. To handle complex instances and cluster customer nodes, a two-stage genetic algorithm is developed. However, the model has some limitations. After deploying one or more robots in customer neighborhoods, the truck can either remain at the drop-off site (Dispatch–Wait–Collect) or return later to retrieve the robots from a predetermined location. This leads to inefficiencies, as the truck is idle during this time and unable to serve additional customers. Furthermore, the model does not account for payload capacity, robot travel distance, or charging features. Xiao et al. \cite{xiao2024cooperative} explored the collaboration between a single truck and a single drone to serve a single customer per trip in rural last-mile delivery scenarios with steep roads. Their model extends the basic constraints of VRP-D models, aiming to minimize total energy consumption during delivery. Hong et al. \cite{hong2024collaborative} investigated the route optimization problem for cooperative meal delivery to elderly individuals, partially synchronizing a single truck with multiple robots. A MILP model is formulated to minimize total system costs, and an enhanced ALNS algorithm is developed for large-scale instances, integrating simulated annealing and artificial bee colony methods to avoid local optima. However, the study has some shortcomings: robots are solely responsible for serving customers in cyclic operations, while key factors such as travel distance, energy consumption, and charging features for robots are not considered. Babaee et al. \cite{tirkolaee2024traveling} modeled the Traveling Salesman Problem with Drones and Bicycles (TSP-D-B), proposing a MILP model to minimize total travel time. While their work highlighted several applications, it lacks a solution method for large-scale instances. Additionally, the model is relatively simple, as only drones and bicycles serve the customers, incorporating partial synchronization among the fleet and addressing the feature of unreachable nodes for trucks and drones. The model does not include features such as multi-visit, multi-trip operations, or charging capabilities.

Compared to research on TSP-D or FSTSP, relatively fewer studies investigate scenarios involving multiple drones collaborating with a fleet of trucks. In the following, studies that consider multiple trucks and multiple drones/robots are discussed.

Jiang et al. \cite{jiang2024multi} studied a multi-visit flexible-docking vehicle routing problem where multiple trucks and drones collaboratively fulfilled pickup and delivery requests in rural areas. Drones can serve multiple customers per trip and dock with the same or different trucks, enabling simultaneous pickup and delivery. The problem is formulated as a MILP model aimed at minimizing total transportation costs and is solved using a customized ALNS metaheuristic. However, the study has several limitations: it does not include an energy consumption model for drones, supports only acyclic operations, lacks recharging features, and omits sensitivity analysis. Rave et al. \cite{rave2023drone} address the two-echelon location routing problem with drones, optimizing the fleet composition of trucks and drones as well as the placement of micro- depots to minimize costs. A MILP formulation is employed and an ALNS is developed to generate near-optimal solutions for larger instances.  However, their model does not account for drones performing multi-trip and multi-visit operations. Additionally, it only considers payload capacity for trucks and excludes energy consumption modeling and recharging capabilities. Wu et al. \cite{wu2022collaborative} investigated a collaborative truck-drone routing problem for contactless parcel delivery. Their model allows multiple trucks and drones to cooperatively deliver parcels, aiming to minimize delivery time while incorporating an energy consumption model for drones. A mixed-integer programming model is proposed, and an enhanced variable neighborhood descent algorithm is developed, combining the Metropolis acceptance criterion from Simulated Annealing and the tabu list. Despite its strengths, their work has some limitations: it does not consider multi-trip operations for drones, lacks recharging features, and does not include experiments to assess the impact of single versus multiple visits. Zhou et al. \cite{zhou2023exact} investigated a two-echelon vehicle routing problem involving collaborative operations between multiple vehicles and drones to serve customers. Drones can perform multiple trips while their paired vehicles remain stationary at customer nodes, creating a two-echelon network. The problem is modeled as a MILP, and a set-partitioning formulation is constructed to improve efficiency. To solve the problem, an exact branch-and-price algorithm is proposed, utilizing a bidirectional labeling algorithm for the pricing subproblem. However, their work has some limitations: only drones are tasked with serving customers, the model does not support flexible docking, and it allows only cyclic operations. Furthermore, their experiments are restricted to small-scale instances with a maximum of $35$ customers, requiring a three-hour time limit to achieve an optimal solution. Yin et al. \cite{yin2023branch} addressed a truck-based drone delivery routing problem with time windows, where each truck operates alongside an associated drone. The drone can launch from the truck at designated nodes, serve one or more customers within their respective time windows, and rejoin the truck at a subsequent node along its route. To solve this problem,  an enhanced branch-and-price-and-cut (BPC) algorithm is proposed, incorporating a bounded bidirectional labeling algorithm to manage the complex pricing subproblem. While the study makes significant contributions, it has some limitations. The synchronization mechanism between trucks and drones is not thoroughly discussed, and the model does not support flexible docking for drones. Additionally, payload capacity and energy consumption constraints are only partially addressed, and recharging options for drones are not considered.

\subsection{Research Gap and Contributions}
Table \ref{tbl_LR} provides a summary of this study's contributions alongside other related works. The comparison evaluates various features, including the number of trucks $(\#\mathcal{T})$, drones $(\#\mathcal{D})$, and robots $(\#\mathcal{R})$, the type of objective, solution methods, collaborative delivery types, operational modes (cyclic or acyclic), handling of unreachable nodes for trucks, fleet synchronization (sync), multi-trip and multi-visit capabilities, and considerations such as maximum distance/flying range, payload capacity, energy consumption modeling, charging features, and the solver used.

\begin{table*}[h]
\centering
\caption{Summary of existing related studies}
\resizebox{\textwidth}{!}{%
\begin{tabular}{|l|lll|l|l|ll|ll|l|c|c|c|c|c|c|c|c|}
\hline
\multirow{2}{*}{\textbf{Ref.}} & \multicolumn{3}{c|}{\textbf{Fleet}}                                                  & \multirow{2}{*}{\textbf{Objective}} & \multirow{2}{*}{\textbf{Solution Method}}& \multicolumn{2}{c|}{\textbf{Collaborative Delivery}} & \multicolumn{2}{c|}{\textbf{Operation}} & \multirow{2}{*}{\textbf{\begin{tabular}[c]{@{}c@{}}Unreachable Nodes\\ for Trucks\end{tabular}}}& \multirow{2}{*}{\textbf{Sync}} & \multirow{2}{*}{\textbf{\begin{tabular}[c]{@{}c@{}}Multi-Trip\\ Service\end{tabular}}} & \multirow{2}{*}{\textbf{\begin{tabular}[c]{@{}c@{}}Multi-Visit\\ Service\end{tabular}}} & \multirow{2}{*}{\textbf{\begin{tabular}[c]{@{}c@{}}Max Dist./\\ Endurance\end{tabular}}} & \multirow{2}{*}{\textbf{\begin{tabular}[c]{@{}c@{}}Payload\\ Capacity\end{tabular}}} & \multirow{2}{*}{\textbf{\begin{tabular}[c]{@{}c@{}}Energy\\ Consumpt.\end{tabular}}} & \multirow{2}{*}{\textbf{\begin{tabular}[c]{@{}c@{}}Charging \\ Feature\end{tabular}}} & \multirow{2}{*}{\textbf{Solver}} \\ \cline{2-4} \cline{7-10}
                     & \multicolumn{1}{l|}{\textbf{$\#\mathcal{T}$}} & \multicolumn{1}{l|}{\textbf{$\#\mathcal{D}$}} & \textbf{$\#\mathcal{R}$} &                                     &                                           & \multicolumn{1}{l|}{\textbf{Category-I}}     & \textbf{Category-II}    & \multicolumn{1}{l|}{\textbf{Cyclic}}  & \textbf{Acyclic}  &                                                                                                  &                                  &                                                                                          &                                                                                         &                                                                                          &                                                                                      &                                                                                      &                                                                                      &                                 \\ \hline \hline

{\rule{0pt}{15pt}}{\cite{meng2024multi}}                        & \multicolumn{1}{c|}{$\mathcal{S}$}                               & \multicolumn{1}{c|}{$\mathcal{S}$}                                 & \cellcolor[HTML]{FD6864}         & \begin{tabular}[c]{@{}l@{}}Total travelling cost of trucks and\\ the total energy cost of   drones\end{tabular}                          & Mixed-integer bilinear program and ALNS                      & \multicolumn{1}{l|}{}                                      & \cellcolor[HTML]{67FD9A}      & \multicolumn{1}{l|}{\cellcolor[HTML]{FD6864}  }                &        \cellcolor[HTML]{67FD9A}      &    \cellcolor[HTML]{67FD9A}                                                                                                                   &   \cellcolor[HTML]{67FD9A}                                                     &     \cellcolor[HTML]{FD6864}                                                        &   \cellcolor[HTML]{67FD9A}                                            &   \cellcolor[HTML]{FD6864}           &   \cellcolor[HTML]{67FD9A}                        &    \cellcolor[HTML]{67FD9A}                                                                                      & \cellcolor[HTML]{FD6864}                                                                        & Gurobi                            \\ \hline

{\rule{0pt}{15pt}}{\cite{jiang2024multi}}                        & \multicolumn{1}{c|}{$\mathcal{M}$}                               & \multicolumn{1}{c|}{$\mathcal{M}$}                                 &                           \cellcolor[HTML]{FD6864}         & Total transportation cost              & MILP and ALNS        & \cellcolor[HTML]{67FD9A}                                        &                  \cellcolor[HTML]{67FD9A}      & \multicolumn{1}{l|}{\cellcolor[HTML]{FD6864} }                &         \cellcolor[HTML]{67FD9A}                               &       \cellcolor[HTML]{67FD9A}                                                       &             \cellcolor[HTML]{67FD9A}                                           &       \cellcolor[HTML]{FD6864}                                                      &   \cellcolor[HTML]{67FD9A}                                   &    \cellcolor[HTML]{67FD9A}                             &   \cellcolor[HTML]{67FD9A}              &   \cellcolor[HTML]{FD6864}                                                                                      & \multicolumn{1}{l|}{\cellcolor[HTML]{FD6864}  }                                                      & Gurobi                            \\ \hline

{\rule{0pt}{15pt}}{\cite{madani2024optimization}}                        & \multicolumn{1}{c|}{$\mathcal{S}$}                               & \multicolumn{1}{c|}{$\mathcal{S}$}                                 &      \cellcolor[HTML]{FD6864}                              & Total traveling cost                                                                                                                     & \begin{tabular}[c]{@{}l@{}}ILP and Variable neighborhood search\\ algorithm\end{tabular}                                                                & \multicolumn{1}{l|}{}                                      &  \cellcolor[HTML]{67FD9A}                                                           & \multicolumn{1}{l|}{\cellcolor[HTML]{68CBD0}}                &                                   \cellcolor[HTML]{68CBD0}    &  \cellcolor[HTML]{FD6864}         &    \cellcolor[HTML]{FD6864}                                                     &  \cellcolor[HTML]{67FD9A}                                                          &    \cellcolor[HTML]{67FD9A}                                                           &       \cellcolor[HTML]{FFCC67}                                                                                                               &                                                    \cellcolor[HTML]{67FD9A}      &                    \cellcolor[HTML]{FD6864}                                                                      & \multicolumn{1}{l|}{\cellcolor[HTML]{FD6864}   }                                                            & CPLEX                             \\ \hline

{\rule{0pt}{15pt}}{\cite{yu2022electric}}                        & \multicolumn{1}{c|}{$\mathcal{S}$}                               & \multicolumn{1}{c|}{\cellcolor[HTML]{FD6864}    }                                  & $\mathcal{S}$                               & Total travel cost                                                                                                                        & MIP and ALNS                                                                                                                                            & \multicolumn{1}{l|}{}                                      &                           \cellcolor[HTML]{67FD9A}                                   & \multicolumn{1}{l|}{\cellcolor[HTML]{FD6864}  }                &    \cellcolor[HTML]{67FD9A}                                     &      \cellcolor[HTML]{FD6864}                                                                                                                  &    \cellcolor[HTML]{67FD9A}                                                           &         \cellcolor[HTML]{68CBD0}                                                     &        \cellcolor[HTML]{67FD9A}                                                                                                                  &    \cellcolor[HTML]{67FD9A}                                                                                                               &    \cellcolor[HTML]{67FD9A}  &    \cellcolor[HTML]{67FD9A}                                                                                                                    &   \cellcolor[HTML]{67FD9A}                                   & CPLEX                             \\ \hline

{\rule{0pt}{15pt}}{\cite{ahn2024operational}}                        & \multicolumn{1}{c|}{$\mathcal{S}$}                               & \multicolumn{1}{c|}{\cellcolor[HTML]{FD6864}    }                                  & $\mathcal{M}$                               & \begin{tabular}[c]{@{}l@{}}Daily operational costs.\end{tabular} & Metaheuristic Genetic Algorithm                                                                                                                         & \multicolumn{1}{l|}{\cellcolor[HTML]{67FD9A} }                                      &                                                            & \multicolumn{1}{l|}{\cellcolor[HTML]{67FD9A} }                &     \cellcolor[HTML]{FD6864}                                   &  \cellcolor[HTML]{FD6864}                                                                                                                      &        \cellcolor[HTML]{FD6864}                                                 &   \cellcolor[HTML]{FD6864}                                                          &         \cellcolor[HTML]{67FD9A}                                                                                                                 &        \cellcolor[HTML]{FD6864}                     &                                                     \cellcolor[HTML]{FD6864}               &         \cellcolor[HTML]{67FD9A}                                                                                & \cellcolor[HTML]{FD6864}                                                                         &                                   \\ \hline

{\rule{0pt}{15pt}}{\cite{xiao2024cooperative}}                        & \multicolumn{1}{c|}{$\mathcal{S}$}                               & \multicolumn{1}{c|}{$\mathcal{S}$}                                 &                   \cellcolor[HTML]{FD6864}               & \begin{tabular}[c]{@{}l@{}}Total energy consumption of truck \\ and drone\end{tabular}                                                   & Adaptive ALNS                                                                                                                                           & \multicolumn{1}{l|}{}                                      &       \cellcolor[HTML]{67FD9A}                                                       & \multicolumn{1}{l|}{\cellcolor[HTML]{FD6864}   }                &    \cellcolor[HTML]{67FD9A}                                     &        \cellcolor[HTML]{FD6864}                       &  \cellcolor[HTML]{67FD9A}                                                        &  \cellcolor[HTML]{67FD9A}                                        &   \cellcolor[HTML]{FD6864}                                                     &         \cellcolor[HTML]{67FD9A}                            & \cellcolor[HTML]{67FD9A}                                                         &   \cellcolor[HTML]{67FD9A}                                                                                     & \cellcolor[HTML]{FD6864}                                                               & CPLEX                             \\ \hline

{\rule{0pt}{15pt}}{\cite{hong2024collaborative}}                        & \multicolumn{1}{c|}{$\mathcal{S}$}                               & \multicolumn{1}{c|}{\cellcolor[HTML]{FD6864} }                                  & $\mathcal{M}$                               & Total cost                                                                                                                               & MIP and ALNS                                                                                                                                            & \multicolumn{1}{l|}{\cellcolor[HTML]{67FD9A} }                                      &                                                          & \multicolumn{1}{l|}{\cellcolor[HTML]{FD6864} }                &     \cellcolor[HTML]{67FD9A}                                  &  \cellcolor[HTML]{FD6864}            &           \cellcolor[HTML]{FFCC67}                                             &   \cellcolor[HTML]{67FD9A}                                                          &  \cellcolor[HTML]{68CBD0}                                            &                \cellcolor[HTML]{FD6864}                                                                                                   &      \cellcolor[HTML]{67FD9A}           &      \cellcolor[HTML]{FD6864}                                                                                   & \cellcolor[HTML]{FD6864}                                                                      & CPLEX                             \\ \hline

{\rule{0pt}{15pt}}{\cite{dai2024two}}                         & \multicolumn{1}{c|}{$\mathcal{S}$}                               & \multicolumn{1}{c|}{$2$}          &     \cellcolor[HTML]{FD6864}                             & \begin{tabular}[c]{@{}l@{}}Total driving cost and\\ delivery time\end{tabular}                                                  & \begin{tabular}[c]{@{}l@{}}Fuzzy C-means algorithm and genetic\\ simulated annealing algorithm\end{tabular}                                             & \multicolumn{1}{l|}{}                                      &  \cellcolor[HTML]{67FD9A}                                                            & \multicolumn{2}{l|}{\cellcolor[HTML]{68CBD0}}      & \cellcolor[HTML]{FD6864}      &                      \cellcolor[HTML]{FFCC67}                                  &                                        \cellcolor[HTML]{FD6864}                      &                          \cellcolor[HTML]{FD6864}      &   \cellcolor[HTML]{67FD9A}      &         \cellcolor[HTML]{67FD9A}                                                         &   \cellcolor[HTML]{FD6864}                                                                                       & \cellcolor[HTML]{FD6864}                                                                       &                                   \\ \hline

{\rule{0pt}{15pt}}{\cite{li2023uav}}                         & \multicolumn{1}{c|}{$\mathcal{S}$}                               & \multicolumn{1}{c|}{$3$}                                 &  \cellcolor[HTML]{FD6864}                                & Total delivery cost                                                                                                             & ILP                                                                                                                                                     & \multicolumn{1}{l|}{\cellcolor[HTML]{67FD9A}   }             &                \cellcolor[HTML]{67FD9A}                                                 & \multicolumn{1}{l|}{\cellcolor[HTML]{FD6864}  }                &    \cellcolor[HTML]{67FD9A}                                    &    \cellcolor[HTML]{FD6864}           &         \cellcolor[HTML]{FD6864}            & \cellcolor[HTML]{FD6864}         &   \cellcolor[HTML]{FD6864}                                    &  \cellcolor[HTML]{67FD9A}    &          \cellcolor[HTML]{67FD9A}                                                  &     \cellcolor[HTML]{FD6864}                                                                                    & \cellcolor[HTML]{FD6864}                                                                      &                                   \\ \hline
{\rule{0pt}{15pt}}{\cite{rave2023drone}}                         & \multicolumn{1}{c|}{$\mathcal{M}$}                               & \multicolumn{1}{c|}{$\mathcal{M}$}                                 & {\cellcolor[HTML]{FD6864}  }                                 & \begin{tabular}[c]{@{}l@{}}Total fixed, variable, \\ synchronization and service costs.\end{tabular}                               & MILP and ALNS                                                                                                                                                  & \multicolumn{1}{l|}{\cellcolor[HTML]{67FD9A}     }                                      &        \cellcolor[HTML]{67FD9A}        & \multicolumn{2}{l|}{\cellcolor[HTML]{67FD9A}     }                                                        &  \cellcolor[HTML]{FD6864}                                    &  \cellcolor[HTML]{67FD9A}                                                       &  \cellcolor[HTML]{FD6864}                                                              &  \cellcolor[HTML]{FD6864}                                                                                                                           &   \cellcolor[HTML]{67FD9A}                                                                                                                 & \cellcolor[HTML]{FFCC67}\begin{tabular}[c]{@{}l@{}}Only for\\truck \end{tabular}  &   \cellcolor[HTML]{FD6864}        & \cellcolor[HTML]{FD6864}       \cellcolor[HTML]{FD6864}                                                                    & CPLEX                             \\ \hline

{\rule{0pt}{15pt}}{\cite{xia2023truck}}                         &    \multicolumn{1}{c|}{$\mathcal{S}$}                                    & \multicolumn{1}{c|}{$\mathcal{S}$}                                        &     \cellcolor[HTML]{FD6864}                             & \begin{tabular}[c]{@{}l@{}}Total cost of the drone weight-related \\ cost, fixed vehicle cost and travel \\ distance cost.\end{tabular}     & MIP                                                                                                                                                     & \multicolumn{1}{l|}{}           &       \cellcolor[HTML]{67FD9A}                                                      & \multicolumn{2}{l|}{\cellcolor[HTML]{68CBD0} }                    &              \cellcolor[HTML]{67FD9A}            &       \cellcolor[HTML]{67FD9A}                                                     &    \cellcolor[HTML]{FD6864}                                                          &        \cellcolor[HTML]{FD6864}                                                                                                                   &      \cellcolor[HTML]{FD6864}                                                                                                               & \cellcolor[HTML]{67FD9A}                                                        &   \cellcolor[HTML]{FD6864}                                                                                        & \cellcolor[HTML]{FD6864}                                                                      & CPLEX                             \\ \hline

{\rule{0pt}{15pt}}{\cite{tirkolaee2024traveling}}                         & \multicolumn{1}{c|}{$\mathcal{S}$}                               & \multicolumn{1}{c|}{$\mathcal{S}$}                                 &          \cellcolor[HTML]{FD6864}                        & Total traveling time                                                                                                                     & MILP                                                                                                                                                    & \multicolumn{1}{l|}{}                                      &  \cellcolor[HTML]{67FD9A}                                                              & \multicolumn{1}{l|}{\cellcolor[HTML]{FD6864}   }                &  \cellcolor[HTML]{67FD9A}                                        &   \cellcolor[HTML]{67FD9A}                &   \cellcolor[HTML]{FFCC67}                                                      &   \cellcolor[HTML]{FD6864}                                                            &   \cellcolor[HTML]{FD6864}                                                                                                                         &    \cellcolor[HTML]{FD6864}                                                                                                                  &                                    \cellcolor[HTML]{FD6864}                        &  \cellcolor[HTML]{FD6864}                                                                                        & \cellcolor[HTML]{FD6864}                                                                        & CPLEX                             \\ \hline

{\rule{0pt}{15pt}}{\cite{weng2023cooperative}}                         & \multicolumn{1}{c|}{$\mathcal{S}$}                               & \multicolumn{1}{c|}{$\mathcal{S}$}                                 &                 \cellcolor[HTML]{FD6864}                  & Completion time of all delivery tasks.                                                                                                   & Hybrid metaheuristic                                                                                                                                    & \multicolumn{1}{l|}{\cellcolor[HTML]{67FD9A}  }                                      &                                                         & \multicolumn{2}{l|}{\cellcolor[HTML]{68CBD0}}                                                        &     \cellcolor[HTML]{FD6864}                                                                                                                    &           \cellcolor[HTML]{FD6864}                                               &   \cellcolor[HTML]{67FD9A}                                                           &     \cellcolor[HTML]{FD6864}               &       \cellcolor[HTML]{67FD9A}    &  \cellcolor[HTML]{67FD9A}                                                              &  \cellcolor[HTML]{FD6864}    & \cellcolor[HTML]{FD6864}                                                                        &                                   \\ \hline

{\rule{0pt}{15pt}}{\cite{madani2024hybrid}}                         & \multicolumn{1}{c|}{$\mathcal{S}$}                               & \multicolumn{1}{c|}{$\mathcal{S}$}                                 &                  \cellcolor[HTML]{FD6864}                & Cost                                                                                                                                     & ILP                                                                                                                                                     & \multicolumn{1}{l|}{}      &       \cellcolor[HTML]{67FD9A}                                                       & \multicolumn{1}{l|}{\cellcolor[HTML]{67FD9A} }                &   \cellcolor[HTML]{67FD9A}                                     &   \cellcolor[HTML]{FD6864}                                                                                     & \cellcolor[HTML]{67FD9A}                                                        &  \cellcolor[HTML]{67FD9A}                                                           &  \cellcolor[HTML]{67FD9A}                                                                                                                        &   \cellcolor[HTML]{67FD9A}                                                                                                                 & \cellcolor[HTML]{67FD9A}                                                          &    \cellcolor[HTML]{FD6864}                                                                                     & \cellcolor[HTML]{FD6864}                                  & CPLEX                             \\ \hline

{\rule{0pt}{15pt}}{\cite{thomas2023collaborative}}                         & \multicolumn{1}{c|}{$\mathcal{S}$}                               & \multicolumn{1}{c|}{$\mathcal{M}$}                                 &   \cellcolor[HTML]{FD6864}                               & Delivery completion time         & \begin{tabular}[c]{@{}l@{}}MILP and relax-and-fix with re-couple\\-refine-and-optimize heuristic approach\end{tabular}                        & \multicolumn{1}{l|}{}                                      &                                     \cellcolor[HTML]{67FD9A}                          & \multicolumn{1}{l|}{\cellcolor[HTML]{68CBD0} }                & \cellcolor[HTML]{68CBD0}                                       &    \cellcolor[HTML]{67FD9A}                                                                                                                    &  \cellcolor[HTML]{67FD9A}                                                       &  \cellcolor[HTML]{FD6864}                                                             &   \cellcolor[HTML]{FD6864}                                                                                                                         &  \cellcolor[HTML]{67FD9A}                                 & \cellcolor[HTML]{67FD9A}                                                           &    \cellcolor[HTML]{FD6864}                                                                                        & \cellcolor[HTML]{FD6864}                                                                         & CPLEX                             \\ \hline

{\rule{0pt}{15pt}}{\cite{wang2019routing}}                         & \multicolumn{1}{c|}{$\mathcal{M}$}                               & \multicolumn{1}{c|}{$\mathcal{S}$}                                 &                   \cellcolor[HTML]{FD6864}                & Total finish time                                                                                                                        & Novel routing and scheduling algorithm                                                                                                                  & \multicolumn{1}{l|}{}                                      &                        \cellcolor[HTML]{67FD9A}                                      & \multicolumn{1}{l|}{\cellcolor[HTML]{FD6864} }                &   \cellcolor[HTML]{67FD9A}                                     &   \cellcolor[HTML]{FD6864}                                                                                                                     &   \cellcolor[HTML]{FD6864}                                                      &  \cellcolor[HTML]{67FD9A}                                                            &      \cellcolor[HTML]{FD6864}                                                                                                                       &   \cellcolor[HTML]{67FD9A}                                                                                                                      & \cellcolor[HTML]{67FD9A}                                                              &   \cellcolor[HTML]{67FD9A}                                                                                          & \cellcolor[HTML]{FD6864}                                                                     &                                   \\ \hline

{\rule{0pt}{15pt}}{\cite{peng2019hybrid}}                        & \multicolumn{1}{c|}{$\mathcal{S}$}                               & \multicolumn{1}{c|}{$\mathcal{M}$}                                 &                             \cellcolor[HTML]{FD6864}      & Delivery time                                                                                                                            & Hybrid Genetic Algorithm                                                                                                                                & \multicolumn{1}{l|}{ \cellcolor[HTML]{67FD9A} }                                      &                                                     & \multicolumn{1}{l|}{ \cellcolor[HTML]{67FD9A} }                & \cellcolor[HTML]{FD6864}                                         &  \cellcolor[HTML]{FD6864}                                                                                                                       &   \cellcolor[HTML]{FD6864}                                                       &   \cellcolor[HTML]{67FD9A}                                                          &  \cellcolor[HTML]{67FD9A}                                                                                                                        &  \cellcolor[HTML]{67FD9A}                  &   \cellcolor[HTML]{FD6864}                 &     \cellcolor[HTML]{FD6864}                                                                                    & \cellcolor[HTML]{FD6864}                                                                  &                                   \\ \hline

{\rule{0pt}{15pt}}{\cite{luo2024collaborative}}                        & \multicolumn{1}{c|}{$\mathcal{S}$}                               & \multicolumn{1}{c|}{$\mathcal{M}$}                                 &              \cellcolor[HTML]{FD6864}                      & \begin{tabular}[c]{@{}l@{}}Weight-sum of truck travel, drone \\ energy consumption and overall time\end{tabular}                          & \begin{tabular}[c]{@{}l@{}}Combined K-Means++ clustering, \\ Nearest neighbor search and Greedy \\ strategies to construct feasible solutions.\end{tabular} & \multicolumn{1}{l|}{}                                      &       \cellcolor[HTML]{67FD9A}                                                 & \multicolumn{1}{l|}{\cellcolor[HTML]{FD6864}  }                &  \cellcolor[HTML]{67FD9A}                                      &   \cellcolor[HTML]{FD6864}                                                                                                                     & \cellcolor[HTML]{67FD9A}                                                             &   \cellcolor[HTML]{67FD9A}                                                               &  \cellcolor[HTML]{67FD9A}                                                                                                                             &  \cellcolor[HTML]{FD6864}                                                                                                                   & \cellcolor[HTML]{67FD9A}                                                         &  \cellcolor[HTML]{67FD9A}                                                                                      & \cellcolor[HTML]{FD6864}                                                                         &                                   \\ \hline

{\rule{0pt}{15pt}}{\cite{mishra2024integrated}}                        & \multicolumn{1}{c|}{$\mathcal{S}$}                               & \multicolumn{1}{c|}{$\mathcal{M}$}                                 &       \cellcolor[HTML]{FD6864}                           & Total integrated cost                                                                                                                  & MINLP                                                                                                                                                   & \multicolumn{1}{l|}{\cellcolor[HTML]{67FD9A}  }                                      &                                                           & \multicolumn{1}{l|}{\cellcolor[HTML]{68CBD0} }                &    \cellcolor[HTML]{68CBD0}                                    &    \cellcolor[HTML]{FD6864}                                                                                                                      &  \cellcolor[HTML]{FD6864}                                                         & \cellcolor[HTML]{67FD9A}                                                           &    \cellcolor[HTML]{67FD9A}                                                                                                                     &  \cellcolor[HTML]{67FD9A}                                                                                                                 &  \cellcolor[HTML]{67FD9A}                                                       &   \cellcolor[HTML]{67FD9A}                                                                                    & \multicolumn{1}{l|}{\cellcolor[HTML]{67FD9A}}                                                                & CPLEX                             \\ \hline

{\rule{0pt}{15pt}}{\cite{wu2022collaborative}}                        & \multicolumn{1}{c|}{$\mathcal{M}$}                               & \multicolumn{1}{c|}{$\mathcal{M}$}                                 &        \cellcolor[HTML]{FD6864}                          & Delivery completion time                                                                                                                 & MIP and Variable Neighborhood Descent                                                                                                                   & \multicolumn{1}{l|}{}                                      &   \cellcolor[HTML]{67FD9A}                                                          & \multicolumn{1}{l|}{\cellcolor[HTML]{FD6864}     }                &      \cellcolor[HTML]{67FD9A}                                 &  \cellcolor[HTML]{FD6864}                                                                                                                      &  \cellcolor[HTML]{67FD9A}                                                        &     \cellcolor[HTML]{FD6864}                                                           &      \cellcolor[HTML]{67FD9A}                                                                                                                       &     \cellcolor[HTML]{67FD9A}                                                                                                                  &       \cellcolor[HTML]{67FD9A}                                                      &   \cellcolor[HTML]{67FD9A}                                                                                        & \cellcolor[HTML]{FD6864}                                                                           & Gurobi                            \\ \hline

{\rule{0pt}{15pt}}{\cite{zhou2023exact}}                        & \multicolumn{1}{c|}{$\mathcal{M}$}                               & \multicolumn{1}{c|}{$\mathcal{M}$}                                 &        \cellcolor[HTML]{FD6864}                          & Total delivery time                                                                                                                 & MILP and Branch-and-price algorithm                                                                                                             & \cellcolor[HTML]{67FD9A}                                     &           &   \cellcolor[HTML]{67FD9A}              &   \cellcolor[HTML]{FD6864}         &       \cellcolor[HTML]{67FD9A}    &        \cellcolor[HTML]{68CBD0}                        &   \cellcolor[HTML]{67FD9A}     &  \cellcolor[HTML]{FD6864}               &      \cellcolor[HTML]{FD6864}                                                  &  \cellcolor[HTML]{67FD9A}                                                     &                                                                             \cellcolor[HTML]{67FD9A}               &   \cellcolor[HTML]{FD6864}                           &  CPLEX                  \\ \hline

{\rule{0pt}{15pt}}{\cite{yin2023branch}}                        & \multicolumn{1}{c|}{$\mathcal{M}$}                               & \multicolumn{1}{c|}{$\mathcal{M}$}                                 &       \cellcolor[HTML]{FD6864}                              &  Total operational cost      &      Arc-based MILP and BPC algorithm                                                                                                    &  \cellcolor[HTML]{67FD9A}                                  &   \cellcolor[HTML]{67FD9A}      &         \cellcolor[HTML]{FD6864}           &     \cellcolor[HTML]{67FD9A}       & \cellcolor[HTML]{67FD9A}              &   \cellcolor[HTML]{68CBD0}                          & \cellcolor[HTML]{67FD9A}          &   \cellcolor[HTML]{67FD9A}           &    \cellcolor[HTML]{FFCC67}                                                &    \cellcolor[HTML]{FFCC67}         &                                           \cellcolor[HTML]{FFCC67}                                                                      &  \cellcolor[HTML]{FD6864}       &  CPLEX                \\ \hline

{\rule{0pt}{15pt}}\textbf{Ours}                 & \multicolumn{1}{c|}{$\mathcal{M}$}                               & \multicolumn{1}{c|}{$\mathcal{M}$}        & $\mathcal{M}$                               &    \begin{tabular}[c]{@{}l@{}}Total operational cost and \\ makespan \end{tabular}                                                                                                                     & MILP and Heuristic Algorithm                                                                                                                                                & \multicolumn{1}{l|}{ \cellcolor[HTML]{67FD9A}   }         &     \cellcolor[HTML]{67FD9A}                                                              &  \cellcolor[HTML]{67FD9A}                                                                 & \multicolumn{1}{l|}{ \cellcolor[HTML]{67FD9A}     }                                      &              \cellcolor[HTML]{67FD9A}                                                                                                               &        \cellcolor[HTML]{67FD9A}                                                      &   \cellcolor[HTML]{67FD9A}                                                               &     \cellcolor[HTML]{67FD9A}                                                                                                                          &       \cellcolor[HTML]{67FD9A}                                                                                                                  &        \cellcolor[HTML]{67FD9A}                                                       &    \cellcolor[HTML]{67FD9A}                                                                                         & \multicolumn{1}{l|}{ \cellcolor[HTML]{67FD9A}     }                                                                    & Gurobi                         \\ \hline \hline
\multicolumn{19}{|l|}{\textbf{Legend:} {\color[HTML]{67FD9A} \textbf{Green:}} Covered, {\color[HTML]{FFCC67} \textbf{Orange:}} Partially Covered, {\color[HTML]{FD6864} \textbf{Red:}} Not Covered, {\color[HTML]{68CBD0} \textbf{Light Blue:}} Not discussed} \\

\multicolumn{19}{|l|}{\textbf{$\mathcal{T}:$} Truck, \textbf{$\mathcal{D}:$} Drone, \textbf{$\mathcal{R}:$} Robot, \textbf{$\mathcal{S}:$} Single, \textbf{$\mathcal{M}:$} Multiple}\\
\hline
\end{tabular}
}
\label{tbl_LR}
\end{table*}
                     
As presented in Table \ref{tbl_LR}, the majority of studies focus primarily on truck-drone or truck-robot collaborations, with only a few \cite{tirkolaee2024traveling}\cite{chen2024multi} integrating bicycles into the delivery system. The studies listed in Table \ref{tbl_LR} overlook the potential benefits of multi-platform combined delivery systems involving three platforms: trucks, drones, and robots. Additionally, only a small number of studies have considered multiple trucks and drones working together; the majority are limited to single-truck single-drone/robot or single-truck multiple-drone/robot scenarios. The concept of multi-visits, where a drone serves multiple customers in a single trip, has also been explored in limited cases \cite{madani2022hybrid}. Notably, only a few studies have addressed the concept of multi-trips allowing drone/robot to make multiple trips. Furthermore, flexible docking, where a drone is not restricted to a dedicated truck has been considered only in a few works \cite{masmoudi2022vehicle, kitjacharoenchai2019multiple, jiang2024multi}. While many studies impose constraints on the maximum flying distance of drones, they often overlook energy consumption constraints, which reduces the optimality of the models \cite{luo2024collaborative}. Moreover, majority of the studies in the literature do not fully leverage the time during which a truck transports a drone or robot for en-route charging and often neglect to consider recharging scenarios altogether, representing a significant gap in the field.

This study addresses these identified gaps in the existing literature by introducing a collaborative multi-platform parcel delivery system for last-mile logistics. The proposed approach enables seamless synchronized collaboration among three platforms: multiple trucks, drones, and robots. The entire fleet can serve the customers, with drones and robots capable of launching from and returning to the truck. Additionally, en-route charging is utilized, maximizing the time the truck carries the drone and robot to enhance system charging efficiency and potentially improve overall performance. The drones and robots can operate in both cyclic and acyclic modes while maintaining tight synchronization with the trucks. They also support flexible docking, multi-visit, and multi-trip operations, provided energy consumption, distance range, and payload capacity constraints are satisfied. This comprehensive approach significantly advances the capabilities of last-mile delivery systems.

\section{Collaborative Multi-platform Last-Mile Delivery}
\label{methodology}

\subsection{Problem Description}
\label{problem_description}
In this research, the VRP-DR models the challenge of last-mile parcel delivery to a set of customers by deploying a fleet of trucks each equipped with multiple drones and robots. The process starts by departing each truck from a single depot, carrying both the parcels and the auxiliary vehicles (drone and robot). Each node (customer) must be visited exactly once, either by a truck, drone, or robot. Once all the parcels are delivered, all trucks, drones, and robots must return to the depot. In this study, both the trucks and auxiliary vehicles can serve the customers simultaneously. Trucks are capable of launching small number of drones and robots, which then carry out the mission sorties to handle nearby deliveries while the truck continues along its designated route to deliver the parcels. 

The structure of a sortie for a drone or robot is represented as a triplet $\left ( i, l, k \right)$, where $i \in \mathcal{V}$ represents the launch node, i.e., the location from which the drone/robot departs from the truck. The set $\mathcal{L}$ contains all possible ordered customer sequences that can be served in a single sortie. Each sequence $l\in\mathcal{L}$ is an ordered tuple of customers from $\mathcal{C}$, constrained by the sortie capacity $m$, which represents the maximum number of customers a drone/robot can serve in one sortie. Formally, this is defined as $ \mathcal{L} \triangleq \left\{ \left ( c_{1},c_{2},c_{3},\cdots,c_{n} \right ) | \ c_{j}\in \mathcal{C}, c_{j}\neq c_{k}, j \neq k,  1\leq n \leq m\right\}$. Finally, $k\in\mathcal{V}$ represents the recovery node, where the drone/robot returns to the truck upon completing the sortie. There is no fixed coupling between the trucks and the auxiliary vehicles, meaning they can launch from and return to different trucks than the ones they initially departed from—a concept referred to as flexible docking. In this study, drones and robots are capable of performing multiple delivery trips referred to as multi-trip and serving multiple customers per trip referred to as multi-visits. This contrasts with most of the literature, which models a single delivery per trip. The drones and robots are only permitted to launch from and return to the truck at a customer node and are not allowed to rendezvous with trucks at intermediate locations. The drones and robots are subject to constraints on maximum payload capacity $\rho_{d}$, $\rho_{r}$ and maximum travel distance range $\mathcal{D}^{d}_{max}$, $\mathcal{D}^{r}_{max}$, respectively. Since drones and robots are limited by battery capacity and cannot serve all customers in a wide area at once, trucks are also employed as mobile charging platforms, expanding the reach of the auxiliary vehicles. When a drone's or robot's battery is depleted, they can return to the truck, receive partial or full charging while the truck is moving, and then continue serving customers.

The proposed problem can be defined as a directed graph $\mathcal{G} = (\mathcal{V}, \mathcal{E})$ where $\mathcal{V} \triangleq  \{ 0,1,2,\cdots , n\}$ is the set of all nodes, with $0$ denoting the depot. $\mathcal{C} \triangleq \mathcal{V} \setminus \{0\}$ represents the set of all customer nodes and each customer $j \in \mathcal{C}$ has a parcel demand measured in weight units $w_{j}$. The edge set $ \mathcal{E} \triangleq \left\{ \left ( i , j \right ): i, j \in \mathcal{V}, i \neq j\right\}$ correspond to the delivery routes between nodes. Each edge $(i,j) \in \mathcal{E}$ is associated with a travel distance for each modality. Each edge also has an associated cost, and the truck, drone, and robot travel at constant speeds $s_{t}, s_{d},$ and $s_{r}$, respectively. Each customer can be served by any vehicle in the fleet. However, to make the model more realistic, certain nodes may not be accessible to trucks due to factors such as narrow streets or restricted areas. In such cases, those nodes are served by either the drone or robot. 

\begin{figure}
    \includegraphics[width=0.99\linewidth]{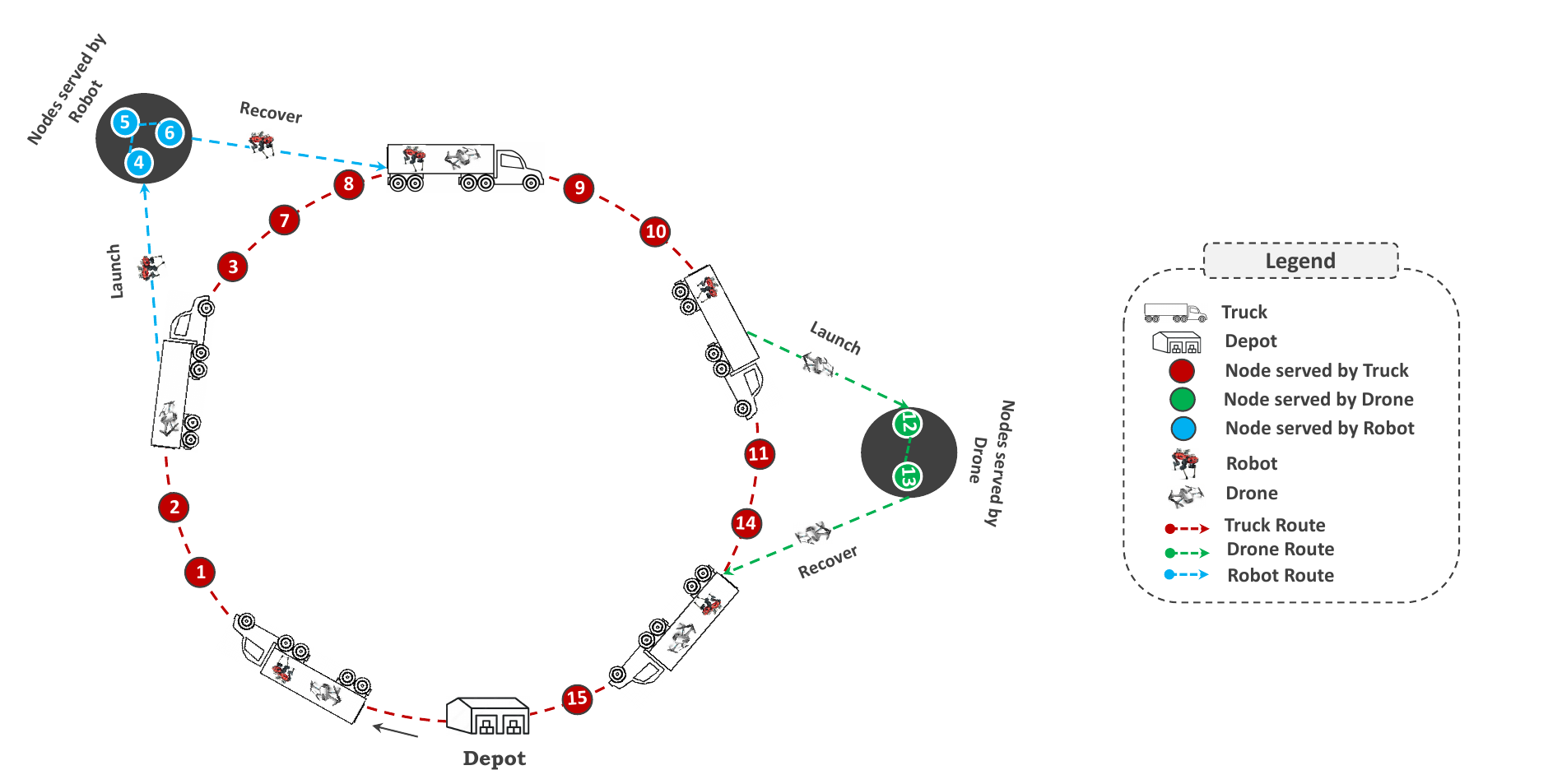}
    \caption{Schematic illustration of the proposed VRP-DR}
    \label{VRP-DR_illustration}
\end{figure}

\begin{figure}
    \includegraphics[width=0.99\linewidth]{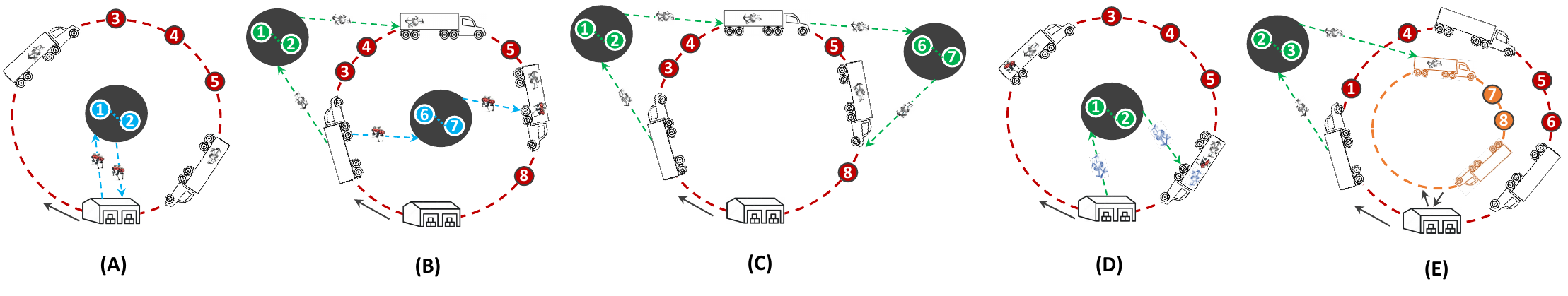}
    \caption{Illustration of multiple scenarios for drone and robot launch and recovery operations}
    \label{launch_and_recover_scenarios}
\end{figure}

To better represent the network, Figure \ref{VRP-DR_illustration} presents a schematic illustration of a simple example of the VRP-DR model with $15$ customers. The truck begins with the trip from the depot, carrying one drone and one robot. It serves customers $1$ and $2$ and launches the robot at customer $3$. The robot serves customers $4$, $5$ and $6$ while the truck continues its route, serving customers $7$ and $8$. The truck stops at customer $9$ to serve them with the robot returning to the truck at customer $9$. The truck then serves customer $10$ and launches the drone at customer $11$, which serves customers $12$ and $13$. The truck continues serving customers $14$ and $15$ with the drone recovering at customer $15$. After serving all the customers collaboratively, the truck, along with all the auxiliary vehicles returns to the depot to complete the trip.
\vspace{0.2pt}
It is also important to highlight that various scenarios for drone and robot launch and recovery operations are considered. Figure \ref{launch_and_recover_scenarios} summarizes these configurations, applicable to both drones and robots, as detailed below:
\begin{itemize}
\item Figure \ref{launch_and_recover_scenarios} (A) illustrates that the robot can directly launch from and return to the depot, represents a cyclic operation.
\item Figure \ref{launch_and_recover_scenarios} (B) presents two scenarios: i) both the drone and robot can launch simultaneously from the same truck at node $3$, and ii) they return to different nodes ($5$ and $8$, respectively), representing an acyclic operation.
\item Figure \ref{launch_and_recover_scenarios} (C) shows a fleet consisting of a single drone and truck, where the drone performs multiple delivery trips. In the first trip, it launches at customer $3$ and serves customers $1$ and $2$, and returns at customer $5$. It launches again from customer $5$, serves customers $6$ and $7$, and then returns to node $8$.
\item Figure \ref{launch_and_recover_scenarios} (D) involves a fleet consisting of two drones, a robot, and a truck. It demonstrates that a second drone (depicted in blue) can directly launch from the depot and return to the truck. 
\item Figure \ref{launch_and_recover_scenarios} (E): this scenario involves a fleet consisting of two trucks and one drone. It illustrates that the drone or robot can launch from one truck and return to a different truck. In this example, the first truck (shown in black) launches the drone at node $1$. The drone serves customers $2$ and $3$ and then returns to node $7$ of the nearest truck $2$ (shown in orange).
\end{itemize}
In addition, the proposed model allows for various types of delivery modes:
\begin{itemize}
   \item \textit{Type I: Truck-Only Delivery---} In this scenario, only the trucks are used to serve all the customers.
   \item \textit{{Type II: Collaborative Delivery---} }In this scenario, either truck-drone or truck-robot can collaboratively be used to serve all the customers
   \item \textit{Type III: Entire Fleet Delivery---} In this scenario, the entire fleet can be used to serve the customers. However, the model optimizes itself to determine which vehicle is best suited to serve each customer. 
\end{itemize}
Concludingly, the VRP-DR problem aims to determine the optimal routes for the trucks and for the drone and robot sorties to collaboratively serve all the customers while minimizing the total operational cost and makespan.
\subsection{Assumptions}
The following assumptions are made in this study:
\begin{itemize}
    \item The network consists of a single depot, designated as node $0$, which serves as the starting point of the truck’s route.
    \item The truck has sufficient capacity to carry both parcels and the auxiliary vehicles.
    \item The speed of each vehicle type is assumed to be known and constant.
    \item The setup time for launching and retrieving drones and robots is disregarded, as it is negligible compared to their travel time.
    \item Each customer has a fixed demand for one parcel, and deliveries are not split.
\end{itemize}
\subsection{Mathematical Formulation}
The VRP-DR is formulated as MILP. The objective function and the constraints associated with the properties of VRP-DR are discussed in detail in this section. The list of notation is provided in Table \ref{tbl_notation}.

The objective of the MILP model in Equation (\ref{Obj_Function}) aims to minimize the weighted sum of the total operational cost and the total operation time (makespan). The operational cost consists of two parts: the fixed cost and the variable cost. The makespan refers to the total time required to complete all deliveries, encompassing the time taken by the truck, drones, and robots to serve all customers. The first component of the objective function represents the unit costs $\mathcal{C}_{t}, \mathcal{C}_{d}, \mathcal{C}_{r}$ associated with the distance traveled by trucks, drones, and robots, respectively. The second component accounts for the fixed costs $f_{t}, f_{d}, f_{r}$ associated with deploying each vehicle type (trucks, drones, and robots), ensuring that the model considers not only operational costs but also the cost of utilizing the vehicles. The third component seeks to minimize the makespan $\Gamma$. The weighting coefficient $\alpha$ is introduced to adjust the sensitivity between operational costs and time efficiency. By tuning the coefficient, the model can prioritize either cost savings or faster deliveries, depending on specific business needs. $x_{ij}^{t}$ is a binary decision variable which is $1$ if truck $t$ travels from node $i$ to $j$, $0$ otherwise. $y_{ilkd}^{t_{i}t_{k}}$ is a binary variable, which is, $1$ if drone $d$ travels from node $i$ to node $k$ serving customer sequence $l$ with truck $t_{i}$ at the launch node and with truck $t_{k}$ at the recovery node, $0$ otherwise. $z_{ilkr}^{t_{i}t_{k}}$ is a binary variable, which is $1$ if robot $r$ travels from node $i$ to node $k$ serving customer sequence $l$ with truck $t_{i}$ at the launch node and with truck $t_{k}$ at the recovery node, $0$ otherwise. \\

\textbf{Objective function:}
\begin{equation}
\label{Obj_Function}
    \begin{aligned}
        \textbf{Minimize} \quad \mathcal{Z} = \alpha \cdot \Big(\sum_{i \in \mathcal{V}} \sum_{j \in \mathcal{V}, i \neq j} \sum_{t \in \mathcal{T}} \mathcal{C}_t \cdot d^t_{ij} \cdot x_{ij}^{t} \ + \\ \sum_{i, k\in \mathcal{V}} \sum_{l \in \mathcal{L}} \sum_{d \in \mathcal{D}} \sum_{t_{i},t_{k} \in \mathcal{T}} \mathcal{C}_d \cdot d^d_{ilk} \cdot y_{ilkd}^{t_{i}t_{k}} \ + \\ \sum_{i,k \in \mathcal{V}} \sum_{l \in \mathcal{L}} \sum_{r \in \mathcal{R}} \sum_{t_{i},t_{k} \in \mathcal{T}} \mathcal{C}_r \cdot d^r_{ilk} \cdot z_{ilkr}^{t_{i}t_{k}} + \\ f_t \cdot \sum_{j \in \mathcal{C}} \sum_{t \in \mathcal{T}} \cdot x_{0j}^{t} \ + f_d \cdot \sum_{i,k \in \mathcal{V}} \sum_{l \in \mathcal{L}} \sum_{d \in \mathcal{D}} \sum_{t_{i},t_{k} \in \mathcal{T}} \cdot y_{ilkd}^{t_{i}t_{k}} \ + \\ f_r \cdot \sum_{i,k \in \mathcal{V}} \sum_{l \in \mathcal{L}} \sum_{r \in \mathcal{R}} \sum_{t_{i},t_{k} \in \mathcal{T}} \cdot z_{ilkr}^{t_{i}t_{k}} \Big) + (1-\alpha) \cdot \Gamma
    \end{aligned}
\end{equation}

\begin{table}[h]
\caption{List of notations and their description}
\centering
\resizebox{\textwidth}{!}{%
\begin{tabular}{ll}
\hline
\multicolumn{1}{l}{{Notation}} & {Description} \\ \hline 
\multicolumn{2}{c}{{Sets}}                            \\ 
\hline\multicolumn{1}{l}{$\mathcal{V}$}                  &  Set of all nodes where $0$ denotes the depot               \\ 
\multicolumn{1}{l}{$\mathcal{C}$}                  &  Set of customer nodes, $\mathcal{C} = \mathcal{V}\setminus \{0\}$          \\ 
\multicolumn{1}{l}{$\mathcal{L}$}                  &   \begin{tabular}[c]{@{}l@{}} Set of all possible ordered customer sequences in a single sortie, where each sequence \\ is constrained by the sortie capacity $m$.
\end{tabular}   \\ 
\multicolumn{1}{l}{$\mathcal{T}$}                  & Set of trucks            \\ 
\multicolumn{1}{l}{$\mathcal{D}$}                  &   Set of drones                 \\ 
\multicolumn{1}{l}{$\mathcal{R}$}                  &   Set of drones                 \\
\hline
\multicolumn{2}{c}{{Parameters}}                      \\ \hline
\multicolumn{1}{l}{$m$}                  &    Maximum customers served per sortie        \\ 
\multicolumn{1}{l}{$d^t_{ij}$}                  &    Manhattan distance between nodes $i$ and $j$ for trucks            \\ 
\multicolumn{1}{l}{$d^d_{ilk}$}                  &   \begin{tabular}[c]{@{}l@{}}Euclidean distance of drone $d$ sortie from node $i$ to node$k$ serving customer sequence $l$ \end{tabular}   \\ 
\multicolumn{1}{l}{$d^r_{ilk}$}                  &   \begin{tabular}[c]{@{}l@{}}Manhattan distance of robot $r$ sortie from node $i$ to node $k$ serving customer sequence $l$ \end{tabular}   \\ 
\multicolumn{1}{l}{$s_t$, $s_d$, $s_r$}                     &    Average speed of truck, drone, and robot respectively       \\ 

\multicolumn{1}{l}{$\mathcal{C}_{t}$, $\mathcal{C}_{d}$, $\mathcal{C}_{r}$}                  &    Unit cost of truck, drone and robot respectively            \\ 
\multicolumn{1}{l}{$f_{t}$, $f_{d}$, $f_{r}$}                  &    Fixed cost of truck, drone and robot respectively            \\ 
\multicolumn{1}{l}{$w_{j}$}                  &    Weight of parcel for customer $j \in \mathcal{C}$            \\ 
\multicolumn{1}{l}{$\mathcal{W}_{d}$, $\mathcal{W}_{r}$}                  &    Self-weight of a drone and robot respectively         \\ 

\multicolumn{1}{l}{$\mathcal{D}_{max}^{d}$, $\mathcal{D}_{max}^{r}$}                  &   \begin{tabular}[c]{@{}l@{}} Maximum distances a drone and a robot can travel in a single sortie \end{tabular} 
\\ 
\multicolumn{1}{l}{$\rho_{d}$, $\rho_{r}$}                  &    Maximum payload capacity of a drone and robot         \\ 
\multicolumn{1}{l}{$\alpha_d$}                  &    Flight energy consumption characteristic of drone 
            \\ 
 \multicolumn{1}{l}{$\mathcal{M}$}                  &    Sufficient large integer  \\ 
 \multicolumn{1}{l}{$\mathcal{B}_{d}$, $\mathcal{B}_{r}$}                  &    Battery capacity of drone and robot respectively \\ 
 \multicolumn{1}{l}{$g$}                  &    Acceleration due to gravity \\ 
 \multicolumn{1}{l}{$l_{len}$}                  &    Leg length of robot \\ 
 \multicolumn{1}{l}{$\mathcal{C}_{rate}^d$, $\mathcal{C}_{rate}^r$}                  &    Charging rate for drone $d$ and robot $r$ respectively\\ 
 \multicolumn{1}{l}{$\alpha$}                  &    Weighting coefficient of objective function\\ \hline 
\multicolumn{2}{c}{{Variables}}              \\ \hline
\multicolumn{1}{l}{$x_{ij}^{t}$}                  &    \begin{tabular}[c]{@{}l@{}}Binary variable, 1 if truck $t$ travels from node $i$ to node $j$, $0$ otherwise  \end{tabular}           \\ 
\multicolumn{1}{l}{$y_{ilkd}^{t_{i}t_{k}}$}                  &    \begin{tabular}[c]{@{}l@{}}Binary variable, 1 if drone $d$ flies from node $i$ to node $k$ serving customer sequence $l$, launched from truck $t_i$ and recovered by truck $t_k$
\end{tabular}           \\ 
\multicolumn{1}{l}{$z_{ilkr}^{t_{i}t_{k}}$}                  &    \begin{tabular}[c]{@{}l@{}}Binary variable, 1 if robot $r$ flies from node $i$ to node $k$ serving customer sequence $l$, launched from truck $t_i$ \\ and recovered by truck $t_k$
\end{tabular}           \\ 
\multicolumn{1}{l}{$\Gamma$}                  &    Makespan (total operation time)    \\ 
\multicolumn{1}{l}{$u_i$,$u_j$}                  &    Continuous variables for sub-tour elimination     \\ 
\multicolumn{1}{l}{$\mathcal{E}^{t_{i}t_{k}}_{ilkd}$}                  &    \begin{tabular}[c]{@{}l@{}}Energy consumption for drone $d$ on sortie from node $i$ to node $k$ via $l$ with truck $t$ \end{tabular}     \\ 
\multicolumn{1}{l}{$\mathcal{E}^{t_{i}t_{k}}_{ilkr}$}                  &    \begin{tabular}[c]{@{}l@{}}Energy consumption for robot $r$ on sortie from node $i$ to node $k$ via $l$ with truck $t$ \end{tabular}    \\ 
\multicolumn{1}{l}{$\mathcal{E}^{t_{i}t_{k}\prime}_{ilkr}$}                  &    \begin{tabular}[c]{@{}l@{}} Linearization variable for robot energy consumption\end{tabular}    \\ 
\multicolumn{1}{l}{$\mathcal{C}^{d}_{vt}$}                  &    Charging amount for drone $d$ at node $v$ with truck $t$    \\ 
\multicolumn{1}{l}{$\mathcal{C}^{r}_{vt}$}                  &    Charging amount for robot $r$ at node $v$ with truck $t$    \\ 
\multicolumn{1}{l}{$\mathcal{C}^{vt}_{time}$}                  &    Charging time at node $v$ for truck $t$   \\ 
\multicolumn{1}{l}{$\mathcal{A}^t_{i}$}                  &    Arrival time of truck $t$ at node $i$ \\ 
\multicolumn{1}{l}{$\Gamma_{ilkd}^{t_{i},t_{k}}$}                  &  \begin{tabular}[c]{@{}l@{}} Launch time of drone $d$ on sortie from node $i$ to node $k$ via $l$  \end{tabular}  \\ 
\multicolumn{1}{l}{$\Gamma_{ilkr}^{t_{i},t_{k}}$}                  &   \begin{tabular}[c]{@{}l@{}} Launch time of robot $r$ on sortie from node $i$ to node $k$ via $l$  \end{tabular} \\ \hline
\end{tabular}
}
\label{tbl_notation}
\end{table}
\textbf{Constraints} \\

In what follows, the constraints of the proposed model are discussed in detail:\\
The makespan $\Gamma$ represents the total time required to complete all deliveries, starting from when the first vehicle departs the depot to when the last vehicle returns. The $\Gamma$ is computed through a set of constraints that ensure it is at least as large as the total travel time of any truck $t$, drone $d$, or robot $r$. $\Gamma$ is subject to the following constraints in the proposed model. The variables $s_t$, $s_d$, and $s_r$ represent the speeds of the truck $t$, drone $d$, and robot $r$, respectively.
\begin{equation}
\Gamma \geq \sum_{i \in \mathcal{V}} \sum_{j \in \mathcal{V}, i \neq j} \frac{d^t_{ij}}{s_t} \cdot x_{ij}^t \quad \forall t \in \mathcal{T}
 \label{Const_mkspan_01}
\end{equation}
\begin{equation}
\Gamma \geq \sum_{i,k \in \mathcal{V}} \sum_{l \in \mathcal{L}} \ \ \frac{d^d_{ilk}}{s_d} \cdot y_{ilkd}^{t_{i}t_{k}} \quad \forall d \in \mathcal{D}, t_{i},t_{k} \in \mathcal{T}
\label{Const_mkspan_02}
\end{equation}
\begin{equation}
\Gamma \geq \sum_{i,k \in \mathcal{V}} \sum_{l \in \mathcal{L}} \ \ \frac{d^r_{ilk}}{s_r} \cdot z_{ilkr}^{t_{i}t_{k}} \quad \forall r \in \mathcal{R}, t_{i},t_{k} \in \mathcal{T}
\label{Const_mkspan_03}
\end{equation}
The constraint in Equation (\ref{Const_visit_once}) below ensures that each node (customer) is visited exactly once by either the truck, drone, or robot.
\begin{equation}
    \begin{aligned}
        \sum_{t \in \mathcal{T}}\sum_{i \in \mathcal{V}}\ \cdot \ x_{ij}^{t} + \sum_{t_{i},t_{k} \in \mathcal{T}}\sum_{d \in \mathcal{D}}\sum_{i,k \in \mathcal{V}}\sum_{l \in \mathcal{L}:j \in l} \cdot\ y_{ilkd}^{t_{i}t_{k}} + \\\sum_{t_{i}t_{k} \in \mathcal{T}}\sum_{r \in \mathcal{R}}\sum_{i,k \in \mathcal{V}}\sum_{l \in \mathcal{L}:j \in l} \cdot\ z_{ilkr}^{t_{i}t_{k}} = 1 \quad \quad \forall j \in \mathcal{C}
    \end{aligned}
\label{Const_visit_once}    
\end{equation}
The constraint in Equation (\ref{Const_depot}) below ensures that each truck starts exactly one trip from the depot and ends exactly one trip at the depot. This guarantees that each truck completes a single tour, preventing multiple departures from or returns to the depot.
\begin{equation}
    \sum_{j \in \mathcal{C}} x_{0j}^{t} = 1, \quad \sum_{i \in \mathcal{C}} x_{i0}^{t} = 1, \quad \forall t \in \mathcal{T}
    \label{Const_depot}
\end{equation}
The constraint in Equation (\ref{Const_flow_conservation}) enforces flow conservation, ensuring that the number of vehicles entering a node equals the number of vehicles leaving it. This guarantees that if a truck enters a node, it must also leave.
\begin{equation}
    \sum_{i \in \mathcal{V}, i \neq j} x_{ij}^{t} = \sum_{i \in \mathcal{V}, i \neq j} x_{ji}^{t}, \quad \forall j \in \mathcal{V}, t \in \mathcal{T}
    \label{Const_flow_conservation}
\end{equation}
The constraint in Equation (\ref{Const_subtour_elimination}) enforces sub-tour elimination in truck routes using the Miller-Tucker-Zemlin (MTZ) formulation \cite{desrochers1991improvements}. It prevents the formation of disconnected cycles that exclude the depot. Here, $u_i$ and $u_j$ are continuous variables representing the relative positions of nodes $i$ and $j$ in the truck route. When $x_{ij}^{t}=1$ (indicating that truck $t$ travels from $i$ to $j$), the constraint ensures $u_{j} \geq u_{i} +1$, maintaing a sequential order that eliminates sub-tours.
\begin{equation}
    u_i - u_j + |\mathcal{C}| \cdot x_{ij}^{t} \leq |\mathcal{C}| - 1, \quad \forall i,j \in \mathcal{C}, i \neq j, t \in \mathcal{T}
    \label{Const_subtour_elimination}
\end{equation}
The constraints below in Equations (\ref{Const_D_SE_T_01}) - (\ref{Const_R_SE_T_02}) ensure that drone and robot sorties remain synchronized with truck movements. These constraints enforce that a drone or robot can only be launched from and recovered at locations visited by a truck, thereby maintaining operational feasibility. Specifically, constraint (\ref{Const_D_SE_T_01}) ensures that a drone sortie $(i,l,k)$ can only be launched from node $i$ using truck $t_{i}$ if that truck actually visits node $i$. Similarly, constraint (\ref{Const_D_SE_T_02}) guarantees that a drone can only return to node $k$ if truck $t_{k}$ is present there, ensuring that when drones complete their sortie, they have a truck available at the rendezvous point for recovery. Constraints (\ref{Const_R_SE_T_01}) and (\ref{Const_R_SE_T_02}) impose the same synchronization requirements for robot operations, ensuring that a robot can only be deployed from a node visited by its assigned truck and can only return to a node where a truck is available for retrieval. These constraints establish operational coordination by linking the sortie variables $y_{ilkd}^{t_{i},t_{k}}$ and $z_{ilkr}^{t_{i},t_{k}}$ with the truck route variables $x_{ji}^{t_{i}}$ and $x_{kj}^{t_{k}}$, thereby preventing infeasible operational scenarios where drones or robots would be deployed from or retrieved at locations without truck support.
\begin{equation}
    \begin{aligned}
     \sum_{l \in \mathcal{L}} \sum_{\substack{k \in \mathcal{V}}} \sum_{d \in \mathcal{D}} y_{ilkd}^{t_{i}t_{k}} \leq \sum_{\substack{j \in \mathcal{V} \\ j \neq i}} x_{ji}^{t_{i}} \quad \forall t_{i},t_{k} \in \mathcal{T}, i \in \mathcal{V}
     \end{aligned}
     \label{Const_D_SE_T_01}
\end{equation}
\begin{equation}
    \begin{aligned}
    \sum_{l \in \mathcal{L}} \sum_{\substack{i \in \mathcal{V}}} \sum_{d \in \mathcal{D}} y_{ilkd}^{t_{i}t_{k}} \leq \sum_{\substack{j \in \mathcal{V} \\ j \neq k}} x_{kj}^{t_{k}} \quad \forall t_{i}t_{k} \in \mathcal{T}, k \in \mathcal{V}
    \end{aligned}
    \label{Const_D_SE_T_02}
\end{equation}
\begin{equation}
   \begin{aligned}
     \sum_{l \in \mathcal{L}} \sum_{\substack{k \in \mathcal{V}}} \sum_{r \in \mathcal{R}}  z_{ilkr}^{t_{i}t_{k}} \leq \sum_{\substack{j \in \mathcal{V} \\ j \neq i}} x_{ji}^{t_{i}} \quad \forall t_{i}t_{k} \in \mathcal{T}, i \in \mathcal{V}
   \end{aligned}
    \label{Const_R_SE_T_01}
\end{equation}
\begin{equation}
    \begin{aligned}
       \sum_{l \in \mathcal{L}} \sum_{\substack{i \in \mathcal{V}}} \sum_{r \in \mathcal{R}} z_{ilkr}^{t_{i}t_{k}} \leq \sum_{\substack{j \in \mathcal{V} \\ j \neq k}} x_{kj}^{t_{k}} \quad \forall t_{i}t_{k} \in \mathcal{T}, k \in \mathcal{V}
    \end{aligned}
    \label{Const_R_SE_T_02}
\end{equation}
Collectively, the model allows drones and robots to make multiple trips, considering the implicit limitations imposed by factors such as energy capacity, charging, and synchronization with trucks.

The constraints in Equations (\ref{const_D_acyclic}) and (\ref{const_R_acyclic}) enforce precedence relationships to maintain the acyclic structure of drone and robot sortie operations, respectively. These constraints use auxiliary variables $u_{i}$ and $u_{k}$ to establish a partial ordering among nodes in the vehicle routing problem. For drone sorties, constraint (\ref{const_D_acyclic}) ensures that if a drone sortie is active $y_{ilkd}^{t_{i}t_{k}}=1$, the recovery node $k$ must be visited by its designated truck $t_{k}$ after the launch node $i$ is visited by truck $t_{i}$. Similarly, constraint (\ref{const_R_acyclic}) enforces the same precedence relationship for robot sorties when a robot sortie is active $z_{ilkr}^{t_{i}t_{k}}=1$.
\begin{equation}
\begin{aligned}
u_k \geq u_i + 1 - |\mathcal{V}| \cdot (1 - y_{ilkd}^{t_{i}t_{k}}), \quad \forall i,k \in \mathcal{V}, l \in \mathcal{L}, d \in \mathcal{D}, t_{i},t_{k} \in \mathcal{T}
\end{aligned}
\label{const_D_acyclic}
\end{equation}
\begin{equation}
\begin{aligned}
u_k \geq u_i + 1 - |\mathcal{V}| \cdot (1 - z_{ilkr}^{t_{i}t_{k}}), \quad \forall i,k \in \mathcal{V}, l \in \mathcal{L}, r \in \mathcal{R}, t_{i},t_{k} \in \mathcal{T}
\end{aligned}
\label{const_R_acyclic}
\end{equation}
The payload constraints in Equation (\ref{const_pyload_drone}) and Equation (\ref{const_pyload_robot}) ensure that the total weight of the packages, denoted by $w_{j}$ carried by a drone or a robot during a sortie does not exceed their their respective maximum payload capacities, $\rho_d$ for drones and $\rho_r$ for robots. For the truck, it is assumed to have sufficient capacity to carry a small number of drones, robots, and parcels, and therefore no explicit capacity constraint is imposed.
\begin{equation}
   \sum_{j\in l} w_j \cdot \ y_{ilkd}^{t_{i}t_{k}} \leq \rho_d, \quad \forall i,k \in \mathcal{V}, l \in \mathcal{L}, d \in \mathcal{D}, t_{i},t_{k} \in \mathcal{T}
    \label{const_pyload_drone}
\end{equation}
\begin{equation}
   \sum_{j\in l} w_j \cdot \ z_{ilkr}^{t_{i}t_{k}} \leq \rho_r, \quad \forall i,k \in \mathcal{V}, l \in \mathcal{L}, r \in \mathcal{R}, t_{i},t_{k} \in \mathcal{T}
    \label{const_pyload_robot}
\end{equation}
The constraints in Equation (\ref{Const_max_dist_drone}) and Equation (\ref{Const_max_dist_robot}) ensure that the drone and the robot do not exceed their maximum allowable travel distances. These constraints account for real-world operational limitations, ensuring that the assigned routes remain feasible within their respective travel range capabilities. For each drone $d \in \mathcal{D}$ and each robot $r \in \mathcal{R}$, the total distance traveled in any sortie $ \left ( i, l, k \right )$ must not exceed the maximum allowable distance:
\begin{equation}
\begin{aligned}
d_{ilk}^{d} \cdot y_{ilkd}^{t_{i}t_{k}} \leq \mathcal{D}_{max}^{d}, \quad \forall i,k \in \mathcal{V},l \in \mathcal{L}, d \in \mathcal{D}, t_{i},t_{k} \in \mathcal{T}
\end{aligned}
    \label{Const_max_dist_drone}
\end{equation}
\begin{equation}
\begin{aligned}
d_{ilk}^{r} \cdot z_{ilkr}^{t_{i}t_{k}} \leq \mathcal{D}_{max}^{r}, \quad \forall i,k \in \mathcal{V},l \in \mathcal{L},r \in \mathcal{R},t_{i},t_{k} \in \mathcal{T}
\end{aligned}
    \label{Const_max_dist_robot}
\end{equation}
where $d_{ilk}^{d}$ and $d_{ilk}^{r}$ represent the total distance of the drone and robot sorties from node $i$ to node $k$ via customer sequence $l$, respectively. $\mathcal{D}_{max}^{d}$ and $\mathcal{D}_{max}^{r}$ denote the maximum distances a drone and a robot can travel in a single sortie.

To compute the energy consumption of a drone sortie, the model is adapted from \cite{luo2024collaborative}. The Equation (\ref{Const_energy1_drone}) calculates the energy consumed by a drone during a sortie. 
\begin{equation}
\begin{aligned}
\mathcal{E}_{ilkd}^{t_{i}t_{k}} = \alpha_{d} \cdot \Bigg[ \Big( \mathcal{W}_{d} + \sum_{j \in l} w_{j} \Big) \cdot d_{i,c_{1}} + \sum_{q=1}^{|l|-1} \Big( \mathcal{W}_{d} + \sum_{\substack{j \in l \setminus \left\{ c_{1}, \cdots, c_{q}\right\}}} w_{j} \Big) \cdot d_{c_{q},c_{q+1}} 
+ \mathcal{W}_{d} \cdot d_{c_{|l|},k} \Bigg]
\end{aligned}
\label{Const_energy1_drone}
\end{equation}
In Equation (\ref{Const_energy1_drone}), $\mathcal{E}_{ilkd}^{t_{i}t_{k}}$ denotes the energy consumed by drone $d$ during its sortie from node $i$ to node $k$ via the customer sequence $l$ using truck $t$. The energy consumption accounts for the varying payload throughout the journey, which decreases as parcels are delivered. Here, $\mathcal{W}_d$ is the self-weight of the drone, $w_{j}$ is the weight of the parcel for customer $j$, and the parameter $\alpha_{d}$ is the flight energy consumption characteristic of drone $d$, influenced by factors such as the maximum power, motor conversion efficiency, and lift ratio. For the first leg of the journey from the launch node $i$ to the first customer $c_{1}$, the drone carries all parcels for the entire customer sequence, with a total weight of $\sum_{j \in l} w_{j}$. At each subsequent leg between customers $c_{q}$ and $c_{q+1}$, the payload decreases as parcels are delivered, leaving only $\sum_{\substack{j \in l \setminus \left\{ c_{1}, \cdots, c_{q}\right\}}} w_{j}$ remaining. Finally, for the return journey from the last customer $c_{|l|}$ to the truck at node $k$, the drone carries only its own weight $\mathcal{W}_d$. This formulation precisely models the drone’s dynamic payload, leading to a more accurate estimation of energy consumption. 

The energy consumption model of the robot is presented in Equations (\ref{Const_energy1_robot}) - (\ref{Const_energy5_robot}). 
\begin{equation}
\begin{aligned}
\mathcal{E}_{ilkr}^{t_{i}t_{k}} = \mathcal{P}_{total}(\mathcal{W}_0) \cdot \frac{d_{i,c_1}}{s_r} + \sum_{q=1}^{|l|-1} \mathcal{P}_{total}(\mathcal{W}_q) \cdot \frac{d_{c_q,c_{q+1}}}{s_r} + \mathcal{P}_{total}(0) \cdot \frac{d_{c_{|l|},k}}{s_r}
\end{aligned}
\label{Const_energy1_robot}
\end{equation}
\begin{equation}
\begin{aligned}
\mathcal{W}_0 = \sum_{j \in l} w_j \quad \quad \mathcal{W}_q = \sum_{j \in l \setminus \{c_1, \ldots, c_q\}} w_j
\end{aligned}
\label{Const_energy2_robot}
\end{equation}
\begin{equation}
\begin{aligned}
\mathcal{P}_{total}(\mathcal{W}) = \mathcal{P}_{mech}(\mathcal{W}) + \mathcal{P}_{elec}(\mathcal{W})
\end{aligned}
\label{Const_energy3_robot}
\end{equation}
\begin{equation}
\begin{aligned}
\mathcal{P}_{mech}(\mathcal{W}) = k_1 \cdot (\mathcal{W}_r + \mathcal{W}) \cdot g \cdot s_r \cdot \left(1 + \frac{g}{2 \cdot l_{leg} \cdot s_r^2}\right)
\end{aligned}
\label{Const_energy4_robot}
\end{equation}
\begin{equation}
\begin{aligned}
\mathcal{P}_{elec}(\mathcal{W}) = k_2 \cdot \mathcal{P}_{mech}(\mathcal{W})
\end{aligned}
\label{Const_energy5_robot}
\end{equation}
Equation (\ref{Const_energy1_robot}) computes the total energy consumed by robot $r$ during a sortie from node $i$ to node $k$ while visiting an ordered sequence of customers $l$ using truck $t$. This equation accounts for the decreasing payload as deliveries are made at each customer location. The model captures the initial leg from the launch point $i$ to the first customer $c_1$  with full payload $\mathcal{W}_0$. The middle term sums the energy used between consecutive customer visits with progressively decreasing payloads $\mathcal{W}_q$. The final term captures the energy required to travel from the last customer $c_{|l|}$ to the rendezvous node $k$ with zero payload, as all parcels have been delivered. The total power consumption $\mathcal{P}_{total}(\mathcal{W})$ in Equation (\ref{Const_energy2_robot}) is the sum of mechanical power $\mathcal{P}_{mech}(\mathcal{W})$ and electrical power $\mathcal{P}_{elec}(\mathcal{W})$. Equation (\ref{Const_energy3_robot}) defines the mechanical power where $\mathcal{\mathcal{W}}_r$ is the self-weight of the robot, $\mathcal{W}$ is the current payload, $g$ is the gravitational acceleration, $s_r$ is the robot's speed, and $l_{leg}$ is the leg length of the robot. The term $(1 + \frac{g}{2 \cdot l_{leg} \cdot s_r^2})$ represents an adjustment factor for the robot's gait, taking into account the interplay between gravity, leg length, and speed. Equation (\ref{Const_energy4_robot}) defines the electrical power, which models electrical losses as proportional to the mechanical power, with $k_2$ being a constant related to the efficiency of the robot's electrical systems. This comprehensive energy model ensures accurate accounting of the varying energy requirements as the robot completes its delivery sequence with progressively lighter loads.

The energy consumption model for robots is a quadratic formulation, reflecting the complex interplay between mechanical and electrical power requirements. However, this quadratic model introduces computational challenges for large-scale optimization. To mitigate this issue, a linearization technique is applied using the big-M method in Equations (\ref{Const_Lin_energy1_robot}) - (\ref{Const_Lin_energy4_robot}), where $\mathcal{M}$ is a sufficiently large constant that facilitates the linear approximation.
\begin{equation}
\begin{aligned}
   \mathcal{E}^{t_{i}t_{k}^\prime}_{ilkr} \leq \mathcal{E}^{t_{i}t_{k}}_{ilkr}, \quad \forall i,k \in \mathcal{V}, i \neq k, l \in \mathcal{L}, r \in \mathcal{R}, t_{i},t_{k} \in \mathcal{T}
\end{aligned}
\label{Const_Lin_energy1_robot}
\end{equation}
\begin{equation}
\begin{aligned}
  \mathcal{E}^{t_{i}t_{k}^\prime}_{ilkr} \leq \mathcal{M} \cdot z_{ilkr}^{t_{i}t_{k}}, \quad \forall i,k \in \mathcal{V}, i \neq k, l \in \mathcal{L}, r \in \mathcal{R}, t_{i},t_{k} \in \mathcal{T}
\end{aligned}
\label{Const_Lin_energy2_robot}
\end{equation}
\begin{equation}
\begin{aligned}
 \mathcal{E}^{t_{i}t_{k}^\prime}_{ilkr} \geq \mathcal{E}^{t_{i}t_{k}}_{ilkr} - \mathcal{M} \cdot (1 - z_{ilkr}^{t_{i}t_{k}}), \quad \forall i,k \in \mathcal{V}, i \neq k, l \in \mathcal{L},  r \in \mathcal{R}, t_{i},t_{k}\in \mathcal{T}
\end{aligned}
\label{Const_Lin_energy3_robot}
\end{equation}
\begin{equation}
\begin{aligned}
  \mathcal{E}^{t_{i}t_{k}^\prime}_{ilkr} \geq 0, \quad \forall i,k \in \mathcal{V}, i \neq k, l \in \mathcal{L}, r \in \mathcal{R}, t_{i},t_{k} \in \mathcal{T}
\end{aligned}
\label{Const_Lin_energy4_robot}
\end{equation}
The VRP-DR also incorporates real-world constraints, where certain nodes may be inaccessible to trucks due to narrow streets, low bridges, or restricted areas. To model these limitations, a distance-based approach is applied in constraint Equation (\ref{const_truck_unreachable_nodes}). This constraint ensures that trucks cannot travel between nodes where the distance $d^t_{ij}$ between node $i$ to node $j$ is set to $\mathcal{M}$, effectively rendering certain nodes or routes unreachable. The value $\mathcal{M}$ represents a sufficiently large number, simulating an infinite distance.
\begin{equation}
    \begin{aligned}
        x_{ij}^{t} = 0, \quad \forall i,j \in \mathcal{V}, t \in \mathcal{T} \quad \text{ where } d^t_{ij} \geq \mathcal{M} 
    \end{aligned}
    \label{const_truck_unreachable_nodes}
\end{equation}
The VRP-DR model introduces an en-route charging feature that allows drones and robots to recharge their batteries partially or fully during the delivery process. This charging takes place at the truck, which serves as a mobile charging station, enabling continuous operations without the need for separate stops. The following mathematical formulation and constraints define this behavior.

Firstly, Equations (\ref{const_fullcharge_depot_drone}) and (\ref{const_fullcharge_depot_robot}) ensure that the drones and robots begin their journey fully charged at the depot. Consequently, the constraints in Equations (\ref{const_nocharge_depot_drone}) and (\ref{const_nocharge_depot_robot}) ensure that no charging occurs when they depart from the depot.
\begin{equation}
    \begin{aligned}
       \sum_{l \in \mathcal{L}, \, k \in \mathcal{V}, \, t_{i}t_{k} \in \mathcal{T}} \mathcal{E}^{t_{i}t_{k}}_{0lkd} \leq \mathcal{B}_d, \quad \forall d \in \mathcal{D}
    \end{aligned}
    \label{const_fullcharge_depot_drone}
\end{equation}
\begin{equation}
    \begin{aligned}
        \sum_{l \in \mathcal{L}, \, k \in \mathcal{V}, \, t_{i}t_{k} \in \mathcal{T}} \mathcal{E}^{t_{i}t_{k}^\prime}_{0lkr} \leq \mathcal{B}_r, \quad \forall r \in \mathcal{R}
    \end{aligned}
    \label{const_fullcharge_depot_robot}
\end{equation}
\begin{equation}
    \mathcal{C}^t_{0d} = 0, \quad \forall d \in \mathcal{D}, t \in \mathcal{T}
    \label{const_nocharge_depot_drone}
\end{equation}
\begin{equation}
    \mathcal{C}^t_{0r} = 0, \quad \forall r \in \mathcal{R}, t \in \mathcal{T}
     \label{const_nocharge_depot_robot}
\end{equation}
The constraints in the Equation (\ref{const_D_ChargeAtTruck}) and (\ref{const_R_ChargeAtTruck}) regulate the charging of drones and robots at truck locations within the collaborative multi-platform parcel delivery system. Equation (\ref{const_D_ChargeAtTruck}) enforces charging conditions for drones, while Equation (\ref{const_R_ChargeAtTruck}) applies to robots. In both cases, charging is only permitted at locations where a truck is present. The term $\mathcal{C}^d_{vt}$ represents the charge supplied to drone $d$ at node $v$ using truck $t$, while $\mathcal{C}^r_{vt}$ denotes the equivalent for robot $r$. The charging rate parameters, $\mathcal{C}_{rate}^d$ for drones and $\mathcal{C}_{rate}^r$ for robots, scale the charging amount based on truck presence. The binary decision variables $x_{iv}^{t}$ and $x_{vi}^{t}$ indicate whether truck $t$ travels to or from node $v$, respectively. Thus, the sum $\sum_{i \in \mathcal{V}, i \neq v} (x_{iv}^{t} + x_{vi}^{t})$ acts as an indicator of truck presence at node $v$, allowing charging only when the truck is available. 
\begin{equation}
    \mathcal{C}^d_{vt} \leq  \mathcal{C}_{rate}^d \ \cdot \sum_{i \in \mathcal{V}, i \neq v} (x_{iv}^{t} + x_{vi}^{t}), \quad \forall  v \in \mathcal{V} \setminus \left\{ 0\right\}, d \in \mathcal{D}, t \in \mathcal{T}
    \label{const_D_ChargeAtTruck}
\end{equation}
\begin{equation}
     \mathcal{C}^r_{vt} \leq  \mathcal{C}_{rate}^r \ \cdot \sum_{i \in \mathcal{V}, i \neq v} (x_{iv}^{t} + x_{vi}^{t}), \quad \forall v \in \mathcal{V} \setminus {0}, r \in \mathcal{R}, t \in \mathcal{T}
    \label{const_R_ChargeAtTruck}
\end{equation}
The constraints in Equations (\ref{const_BC_drone}) and (\ref{const_BC_robot}) ensure that the cumulative energy consumed by each drone $d$ and robot $r$ across all sorties does not exceed their respective battery capacities $\mathcal{B}_{d}$, $\mathcal{B}_{r}$, respectively, plus any energy replenished through en-route charging. Here $\mathcal{C}^d_{vt}$ and $\mathcal{C}^r_{vt}$ represent the amount of energy charged by drone $d$ and robot $r$ at node $v$ while onboard truck $t$.
\begin{equation}
   \sum_{i,k \in \mathcal{V}}\sum_{l \in \mathcal{L}}\sum_{t_{i}t_{k} \in \mathcal{T}} \mathcal{E}^{t_{i}t_{k}}_{ilkd} \leq \mathcal{B}_d + \sum_{v \in \mathcal{V} \setminus \{0\}} \mathcal{C}^d_{vt}, \quad \forall d \in \mathcal{D}
    \label{const_BC_drone}
\end{equation}
\begin{equation}
\sum_{i,k \in \mathcal{V}}\sum_{l \in \mathcal{L}}\sum_{t_{i},t_{k} \in \mathcal{T}} \mathcal{E}^{t_{i}t_{k}^\prime}_{ilkr} \leq \mathcal{B}_r + \sum_{v \in \mathcal{V} \setminus \{0\}} \mathcal{C}^r_{vt}, \quad \forall r \in \mathcal{R}
    \label{const_BC_robot}
\end{equation}
The charging time for drone and robot at any node is bounded by the truck's travel time to its next destination in the route, as formulated in Equation (\ref{const_chargingTime}). In this Equation, $\mathcal{C}_{time}^{vt}$ represents the available charging duration at node $v$ for truck $t$, $\frac{d^t_{vi}}{s_t}$ is the travel time from node $v$ to node $i$, and $x_{vi}^{t}$ is the binary decision variable indicating whether truck $t$ travels directly from node $v$ to node $i$. This constraint ensures that auxiliary vehicles can be charged while being transported on the truck, utilizing the travel time between customer locations. The available charging duration is determined by the truck’s movement, allowing for opportunistic energy replenishment without requiring dedicated charging stops.
\begin{equation}
    \mathcal{C}_{time}^{vt} \leq \sum_{i \in \mathcal{V}, i \neq v} \frac{d^t_{vi}}{s_t} \cdot x_{vi}^{t} ,\quad \forall v \in \mathcal{V}, t \in \mathcal{T}
    \label{const_chargingTime}
\end{equation}
Constraints in Equation (\ref{const_D_ChargeLimit}) and (\ref{const_R_ChargeLimit}) limit the amount of charging based on the charging rate and the time spent at each node. 
\begin{equation}
    \mathcal{C}^d_{vt} \leq \mathcal{C}_{rate}^d \cdot \ \mathcal{C}_{time}^{vt}, \quad \forall v \in \mathcal{V}, d\in \mathcal{D}, t \in \mathcal{T}
    \label{const_D_ChargeLimit}
\end{equation}
\begin{equation}
    \mathcal{C}^r_{vt} \leq \mathcal{C}_{rate}^r \cdot \  \mathcal{C}_{time}^{vt}, \quad \forall v \in \mathcal{V}, r \in \mathcal{R}, t \in \mathcal{T}
    \label{const_R_ChargeLimit}
\end{equation}
The constraints in Equations (\ref{const_D_PreventOvercharging}) and (\ref{const_R_PreventOvercharging}) ensure that the total charge of drones and robots does not exceed their respective battery capacities at any point in the route. These constraints monitor the battery state throughout the delivery process by balancing the initial charge, energy consumption, and recharging activities.

For each node $v$ in the route, excluding the depot, the constraint calculates the current battery level by taking the initial capacity $B_{d}$ or $B_{r}$, subtracting the cumulative energy consumed up to node $v$, and adding the cumulative recharging performed up to node $v$, ($\mathcal{C}_{ut}^{d}$ for drones and $\mathcal{C}_{ut}^{r}$ for robots). The resulting charge level must not exceed the maximum battery capacity. Energy consumption is tracked for all sorties ending at or before $v$, while charging is accounted for at all non-depot nodes visited up to $v$.
\begin{equation}
\begin{aligned}
    \mathcal{B}_d \ - \sum_{\substack{i,k \in \mathcal{V}, l \in \mathcal{L}, t_i, t_k \in \mathcal{T} \\ i \neq k, k \leq v}} \mathcal{E}^{t_i t_k}_{ilkd} 
    + \sum_{\substack{u \in \mathcal{V}, t \in \mathcal{T} \\ u \neq 0, u \leq v}} \mathcal{C}_{ut}^{d} 
    \leq \mathcal{B}_d, \forall d \in \mathcal{D}, v \in \mathcal{V} \setminus \{0\}
    \end{aligned}
    \label{const_D_PreventOvercharging}
\end{equation}
\begin{equation}
\begin{aligned}
    \mathcal{B}_{r} \ - \sum_{\substack{i,k \in \mathcal{V}, l \in \mathcal{L}, t_i, t_k \in \mathcal{T} \\ i \neq k, k \leq v}} \mathcal{E}^{t_i t_k^\prime}_{ilkr} 
    + \sum_{\substack{u \in \mathcal{V}, t \in \mathcal{T} \\ u \neq 0, u \leq v}} \mathcal{C}_{ut}^{r} 
    \leq \mathcal{B}_r, \forall r \in \mathcal{R}, v \in \mathcal{V} \setminus \{0\}
    \end{aligned}
    \label{const_R_PreventOvercharging}
\end{equation}
By enforcing these constraints, the model ensures that drones and robots are never overcharged beyond their battery capacities, while still allowing for recharging during the route to extend their operational range.

The constraint in Equation (\ref{const_synch_1}) enforces the temporal sequencing of truck movements within the routing plan. It ensures that if truck $t$ travels directly from node $i$ to node $j$, then the arrival time at node $j$, represented by $\mathcal{A}^t_{j}$, must be at least the arrival time at node $i$, plus the travel time between the two nodes. 
\begin{equation}
         \mathcal{A}^t_{j} \geq \mathcal{A}^t_{i}+ \frac{d_{ij}^{t}}{s_{t}} \cdot x_{ij}^{t} \quad \forall i,j \in \mathcal{V}, i \neq j, t \in \mathcal{T}
        \label{const_synch_1}
\end{equation}
The synchronization constraints in Equation (\ref{const_launch_drone}) - (\ref{const_return_robot}) ensure that the timing of truck movements aligns with the operations of drones and robots. They enforce temporal synchronization between truck and auxiliary vehicles sorties. The constraints in Equation (\ref{const_launch_drone}) and Equation (\ref{const_launch_robot}) govern the synchronization of launch times, while the constraints in Equation (\ref{const_launch_drone}) and Equation (\ref{const_launch_robot}) govern the synchronization of return times for drone and robot deployments.

In Equation (\ref{const_launch_drone}), when a drone sortie is activated $y_{ilkd}^{t_{i}t_{k}}=1$, the launch time of drone $d$ for customer sequence $l$ departing from node $i$ to node $k$$, \Gamma_{ilkd}^{t_{i}t_{k}}$ must be at least equal to the arrival time of truck ${t_{i}}$ at node $i$, $\mathcal{A}^{t_{i}}_{i}$. This constraint ensures that a drone can only depart after its carrier truck has arrived at the launch location. 
\begin{equation}
 \begin{aligned}
    \Gamma_{ilkd}^{t_{i}t_{k}}\geq \mathcal{A}^{t_{i}}_{i} - \mathcal{M} \cdot \left ( 1 - y_{ilkd}^{t_{i}t_{k}} \right ), \quad i,k \in \mathcal{V}, i \neq k, l \in \mathcal{L}, d \in \mathcal{D}, t_{i},t_{k} \in \mathcal{T}
    \end{aligned}
    \label{const_launch_drone}
\end{equation}
Similarly, the constraint in Equation  (\ref{const_launch_robot}) guarantees that when a robot sortie is active $z_{ilkr}^{t_{i}t_{k}} = 1$, the launch time of robot $r$ for customer sequence $l$ traveling from node $i$ to node $k$, $\Gamma_{ilkr}^{t_{i}t_{k}}$ must be greater than or equal to the arrival time of truck $t_{i}$ at node $i$, $\mathcal{A}^{t_{i}}_{i}$. This condition prevents robots from departing before their carrier truck has reached the deployment location.
\begin{equation}
    \begin{aligned}
        \Gamma_{ilkr}^{t_{i}t_{k}}\geq \mathcal{A}^{t_{i}}_{i} - \mathcal{M} \cdot \left ( 1 - z_{ilkr}^{t_{i}t_{k}} \right ), \quad i,k \in \mathcal{V}, i \neq k, l \in \mathcal{L}, r \in \mathcal{R}, t_{i},t_{k} \in \mathcal{T}
    \end{aligned}
    \label{const_launch_robot}
\end{equation}
The constraints in Equation (\ref{const_return_drone}) and Equation (\ref{const_return_robot}) ensure that a drone or a robot launched from location $i$, serving a customer sequence $l$, and returning to location $k$ arrives before the truck $t_{k}$ arrives at the return location $k$. The constraint ensures that if a sortie is selected, then the drone's or robot's departure time plus its travel time must be less than or equal to the truck's arrival time $\mathcal{A}^{t_{k}}_{k}$ at the return location. This synchronizes the auxiliary vehicles and truck operations, ensuring the truck is there to pick up the drone and robot when it completes its deliveries.
\begin{equation}
    \begin{aligned}
        \Gamma_{ilkd}^{t_{i}t_{k}} + \frac{d^{d}_{ilk}}{s_{d}}\leq \mathcal{A}^{t_{k}}_{k} + \mathcal{M}\cdot \left ( 1- y_{ilkd}^{t_{i}t_{k}} \right )\quad i,k \in \mathcal{V}, i \neq k, l \in \mathcal{L}, d \in \mathcal{D}, t_{i},t_{k} \in \mathcal{T}
    \end{aligned}
    \label{const_return_drone}
\end{equation}
\begin{equation}
    \begin{aligned}
         \Gamma_{ilkr}^{t_{i}t_{k}} + \frac{d^{d}_{ilr}}{s_{r}}\leq \mathcal{A}^{t_{k}}_{k} + \mathcal{M}\cdot \left ( 1- z_{ilkr}^{t_{i}t_{k}} \right )\quad i,k \in \mathcal{V}, i \neq k, l \in \mathcal{L}, r \in \mathcal{R}, t_{i},t_{k} \in \mathcal{T}
    \end{aligned}
     \label{const_return_robot}
\end{equation}
The proposed VRP-DR model incorporates a flexible docking mechanism, where drones and robots are not restricted to returning to the same truck from which they launched. This is implemented through decision variables that explicitly account for both the launch truck $t_i$ and return truck $t_k$, along with synchronization constraints that ensure proper timing coordination. The time synchronization constraints ensure proper coordination between the trucks, with drones/robots only launching after their designated truck arrives at the launch node and returning before their designated truck departs from the return node. This flexibility enhances the model's capability to optimize delivery operations across the entire fleet, potentially reducing overall delivery time and transportation costs compared to single-truck docking approaches.

\section{Solution Method}
\label{solution_method}
The VRP-DR problem introduced in this work is NP-hard, making it computationally intractable for large-scale instances. While the MILP formulation guarantees exact solutions for small problem sizes, its computational complexity grows exponentially with problem size. Consequently, solving large-scale instances within a reasonable timeframe becomes infeasible, necessitating the development of an efficient and scalable heuristic approach.

This section introduces FINDER (Flexible INtegrated
Delivery with Energy Recharge), a hierarchical heuristic designed for collaborative multi-platform last-mile delivery. FINDER employs a structured three-phase sequential approach: (1) truck routes are generated using a clustering-based construction method, (2) followed by the synchronized assignment of drones and robots based on the truck timeline, and (3) the insertion of remaining customers, incorporating en-route charging capabilities for drones and robots during parcel delivery. This decomposition ensures computational tractability while maintaining coordination and energy constraints inherent to the VRP-DR. 

The following subsections provide a detailed discussion of each component of the heuristic solution.

\subsection{Route Construction for Trucks}
The FINDER initiates the routing process by constructing initial truck routes using a sequential insertion algorithm. The trucks serve as mobile depots and charging stations for the auxiliary vehicles (drones and robots). The truck route generation process consists of the following key steps, which are presented in Algorithm 1.
\begin{enumerate}
\item \textit{Initial Route Sizing:} The algorithm determines the optimal customer distribution per truck using an adaptive allocation formula $\max (3, \lfloor |\mathcal{C}| / (2\mathcal{T}) \rfloor)$. This ensures balanced route sizing while maintaining operational efficiency through a minimum route length constraint.
\item \textit{Sequential Route Construction:} The route construction follows a parallel insertion strategy:
\begin{itemize}
    \item Multiple empty routes are initialized simultaneously, one for each available truck.
    \item Each route is initialized with the depot $(0)$ as the starting point.
    \item  For each route, the algorithm selects the nearest unassigned customer $j$ based on Euclidean distance $\sqrt{(x_i - x_j)^2 + (y_i - y_j)^2}$ from the last visited location.
    \item The selected customer is simultaneously removed from both the available and unserved customer sets.
    \item The process continues until the route reaches the predefined customer limit per truck, as determined by the adaptive allocation formula.
\end{itemize}
\item \textit{Route Completion:} The route completion follows the following steps:
\begin{itemize}
    \item Once a route reaches its capacity limit, it is closed by adding a return trip to the depot.
    \item The algorithm moves to constructing the next truck's route while considering the remaining unserved customers.
    \item Any partially filled final route is closed with a return to depot, ensuring that all routes form complete cycles.
    \item The process continues until all customers are assigned or all trucks are utilized.
\end{itemize}
\end{enumerate}
This approach prioritizes computational efficiency while maintaining solution quality through balanced customer allocation and distance-based insertion.

\subsection{Temporal Synchronization and Timeline}
Once the truck routes are established, the algorithm constructs a detailed temporal framework to track each truck’s movement, capturing its precise location and corresponding timestamp throughout the entire route. The timeline is generated by iterating through each truck route and computing travel times between consecutive nodes based on their distance and the truck's speed. The timeline is structured as a list of tuples, where each tuple contains the node's identifier and its associated timestamp. This granular tracking of truck positions enables the system to determine the truck's exact location at any given moment. Temporal precision is essential for synchronizing auxiliary vehicle operations, as it provides accurate time windows for potential drone and robot routes. The generated timeline serves multiple strategic purposes. First, it identifies precise truck locations and timestamps that can be leveraged for drone and robot charging. Second, it ensures feasibility by verifying that auxiliary vehicle routes align with truck movements while adhering to sequencing constraints. Lastly, it acts as a temporal coordinate system, allowing drones and robots to be strategically dispatched along truck routes, maximizing resource utilization and operational efficiency.
\begin{algorithm}
\caption{Truck Route Construction for VRP-DR}
\label{alg:truck_routes}
\textbf{Input:} Customer set \(\mathcal{C}\), Depot location \(D\), Number of trucks \(\mathcal{T}\) \\
\textbf{Output:} Truck routes \(R\)
\begin{algorithmic}[1]
\State \textbf{Initialize:}
\State \(R \gets \emptyset\) \Comment{Initialize routes for \(T\) trucks}
\State \(A \gets \mathcal{C}\) \Comment{Available customers set}
\State \(\mathcal{N} \gets \max(3, \lfloor |\mathcal{C}| / (2\mathcal{T}) \rfloor)\) \Comment{Customers per truck}
\State \(t \gets 0\) \Comment{Current truck index}
\While{\(A \neq \emptyset\) \textbf{and} \(t < T\)}
    \State \(R_t \gets [D]\) \Comment{Start new route from depot}
    \While{\(|R_t| - 1 < \mathcal{N}\) \textbf{and} \(A \neq \emptyset\)}
        \State \(p_{\text{current}} \gets R_t[\text{end}]\) \Comment{Current position of truck}
        \State \(c_{\text{nearest}} \gets \arg \min_{c \in A} \text{distance}(p_{\text{current}}, c)\) \Comment{Find nearest customer}
        \State \(R_t \gets R_t \cup \{c_{\text{nearest}}\}\)
        \State \(A \gets A \setminus \{c_{\text{nearest}}\}\)
    \EndWhile
    \State \(R_t \gets R_t \cup \{D\}\) \Comment{Return to depot}
    \State \(t \gets t + 1\)
\EndWhile
\State \Return \(R\)
\end{algorithmic}
\end{algorithm}
\subsection{Auxiliary Vehicle Route Assignment and Synchronization}
The assignment of auxiliary vehicle routes is integrated with the truck route timeline, ensuring synchronization while adhering to multiple operational constraints. The process follows three key steps:
\begin{enumerate}
    \item \textit{Timeline Traversal:} The algorithm iterates through consecutive truck positions in chronological order. At each position, it evaluates potential insertion points for drone and robot routes between consecutive truck stops. This method identifies opportunities to dispatch auxiliary vehicles to serve unassigned customers while maintaining synchronization with truck routes.
    \item \textit{Route Feasibility Check:} Before assigning routes, the algorithm performs comprehensive feasibility checks to ensure operational viability:
    \begin{itemize}
        \item Verifies that customer payloads do not exceed the vehicle’s capacity.
        \item  Ensures that route distances fall within the vehicle’s maximum operational range.
        \item Dynamically evaluates energy consumption and identifies potential charging opportunities along truck routes.
        \item Ensures that the auxiliary vehicle’s route remains synchronized with the truck’s movements.
    \end{itemize}
    \item \textit{Synchronization Requirements:} The route assignment process adheres to strict synchronization requirements:
    \begin{itemize}
        \item An auxiliary vehicle can only launch from a node after the truck has arrived.  
        \item It must complete its route and rendezvous with the truck before the truck departs.
        \item All operations must align precisely with the truck’s timeline to ensure seamless coordination.
    \end{itemize}
\end{enumerate}
The algorithm maintains operational integrity through continuous tracking of vehicle states, customer assignments, and energy levels, with updates reflected in the vehicle availability and charging history records.

\subsection{Energy Management and En-Route Charging}
En-route charging enables auxiliary vehicles to recharge their batteries while being transported on the truck between delivery nodes. This process maximizes operational efficiency by utilizing idle transit time and leveraging direct access to the truck’s power source for faster charging. To optimize energy management, the system dynamically calculates the energy gained during each charging session by multiplying the charging rate by the duration the vehicle remains on the truck. The charging period is determined based on the time elapsed between the truck’s departure from the initial node and its arrival at the destination. To prevent overcharging, a capacity constraint ensures that the total charge does not exceed the vehicle’s maximum battery capacity. Additionally, a detailed charging log is maintained for each vehicle, recording key parameters such as the time and location of charging events and the amount of energy transferred. This charging mechanism is seamlessly integrated into the route planning process, where energy constraints are evaluated to determine route feasibility. Battery levels are updated after each delivery cycle, accounting for energy consumption during operations and incorporating any energy replenished through en-route charging. This continuous tracking ensures accurate battery management, allowing vehicles to maintain sufficient energy levels for their assigned routes while effectively utilizing charging opportunities during truck transport.

\subsection{Unserved Customer Route Insertion}
To incorporate any remaining unserved customers into existing truck routes, the algorithm employs a systematic insertion heuristic. For each unserved customer, the algorithm evaluates the incremental distance cost across all possible insertion points within the truck routes. This cost is computed using a distance differential equation that compares the original segment length with the modified distance after inserting the customer. The differential distance is calculated as $\Delta d = d(i,c) + d(c,i+1) - d(i,i+1)$, where $d(i,j)$ represents the Euclidean distance between node $i$ and node $j$, $c$ denotes the unserved customer to be inserted, $i$ and $i+1$ are two consecutive nodes in the truck route. The optimal insertion position is determined by identifying the minimum distance differential across all potential positions and truck routes, formally expressed as: $(t, pos) = \arg \min\limits_{t, pos}\ \Delta d(t, pos)$. Once the optimal position is identified, the customer is immediately inserted into the selected truck route and removed from the unserved customer set. This process is repeated iteratively until all remaining customers are integrated into the truck routes.

\section{Numerical Experiments}
\label{experimental_results}
This section presents the numerical setup, followed by an evaluation of the model’s effectiveness and performance through a series of numerical experiments covering various aspects of the problem. The mathematical programming model and heuristic solution are implemented in Python v3.12.4, with the mathematical programming model solved using the Gurobi solver v10.0.0 \cite{TheLeade68:online}. All experiments are conducted on a high-performance computing system equipped with an Intel® Xeon® Gold 6230R processor running at $2.1$ GHz and operating on Ubuntu.

\subsection{Instance Generation and Parameter Settings}
The test instances are generated with a single depot serving as the origin for all deliveries. Customer locations are randomly distributed within a rectangular area of $15 \times 15$ $km^2$, following a uniform distribution. This approach, widely adopted in the literature, ensures the representation of diverse spatial configurations \cite{thomas2024collaborative}. To account for different travel modes, distances are computed using the Manhattan metric for truck and robot travel, while the Euclidean metric is applied for drone travel. Each customer’s package weight is randomly generated within the range of $0.5$ to $10$ kg. Each instance is labeled as [Instance Type]-[Number of Nodes]. The parameter values, commonly adopted in the literature and detailed in Table \ref{tbl_parameter_values}, are used to solve the model.
\begin{table}[h]
\centering
\caption{Parameters and their values}
\resizebox{0.51\textwidth}{!}{%
\begin{tabular}{clcc}
\hline
{Parameter} & {Description} & {Unit} & {Value} \\ \hline
   $s_{t}$               & Speed of truck                    &   Km/h            &  45             \\ 
   $s_{d}$               & Speed of drone                    &   Km/h           &    75           \\ 
   $s_{r}$               & Speed of robot              &   Km/h         &    25           \\ 
    $\mathcal{C}_t$               &  Unit cost of truck                    &   \$            &     2.9          \\ 
                  $\mathcal{C}_d$               &  Unit cost of drone                 &   $\$$        &  0.08            \\ 
                  
$\mathcal{C}_r$               &  Unit cost of robot            &   $\$$           &   0.06       \\ 

$f_t$               &  Fixed cost of truck          &  $\$$            &  30          \\ 

$f_d$               & Fixed cost of drone           &   $\$$           &  10          \\ 

$f_r$               &  Fixed cost of robot            &   $\$$            &  8          \\ 

$\rho_d$               &  Payload capacity of drone         &   Kg        &   25         \\ 

$\rho_r$               &  Payload capacity of robot         &   Kg           &  20          \\ 
$w$               &  Weight of package        &   Kg            & [0.5, 10]         \\ 
$\mathcal{D}_{max}^d$             &  \begin{tabular}[c]{@{}l@{}} Maximum distance a drone \\ can travel in one sortie \end{tabular}    &   Km            &  20          \\ 
$\mathcal{D}_{max}^r$             &  \begin{tabular}[c]{@{}l@{}}Maximum distance a robot \\ can travel in one sortie \end{tabular}     &  Km          &   15         \\ 
$\mathcal{W}_d$             &  Self-weight of a drone    &   Kg            &    18        \\ 
$\mathcal{W}_r$             &  Self-weight of a robot    &   Kg          &    15        \\ 
$\mathcal{B}_d$             &  Battery capacity of drone     &   mAh             &    14000       
\\ 
$\mathcal{B}_r$             &  Battery capacity of robot    &    mAh        &   8000         \\ 
$\alpha_d$             & \begin{tabular}[c]{@{}l@{}} Drone energy consumption\\ co-efficient \end{tabular}   &              &    128        \\ 
$g$             &  Acceleration due to gravity    &              &   9.81         \\ 
$l_{leg}$             & Leg length of robot   &      m        &    0.5        \\ 
$C_{rate}^d$             & Charging rate for drone  &      mAh        &   5000         \\ 
$C_{rate}^r$             & Charging rate for robot  &      mAh        &   4000        \\ 
$k_1$, $k_2$             &  \begin{tabular}[c]{@{}l@{}}Constants related to robot \\ energy consumption \end{tabular} &           &   0.1, 0.2         \\ 
$\alpha$      & \begin{tabular}[c]{@{}l@{}} Objective function \\ weighting coefficient \end{tabular}  &              &   0.5    \\ \hline        
\end{tabular}
}
\label{tbl_parameter_values}
\end{table}
\subsection{Performance Analysis}
This section evaluates the effectiveness of the proposed exact (MILP) and heuristic solutions through a series of experiments on instances of varying scales: small (S), medium (M), and large (L). The experiments are conducted with a fleet consisting of one truck carrying one drone and one robot. It is assumed that all customers are accessible and can be served by either vehicle. The computational results are summarized in Tables \ref{Small_Medium_Instances_Obj}, \ref{Small_Medium_Instances_Time}, \ref{Large_Instances_Obj}, and \ref{Large_Instances_Time}, providing a comparative analysis of the solutions in terms of objective value (solution quality) and computational time. For the small and medium-sized instances, the MIP gap in Gurobi is set to $0\%$, and no time limit is applied to ensure the optimal solution is found and to assess the solver's computational speed. To ensure statistical robustness, each experiment is repeated $25$ times, and the average and standard deviation of the results are reported for both the exact and heuristic solutions. This analysis offers a comprehensive assessment of the performance and efficiency of the proposed approaches across different problem scales.

\subsubsection{Experiment with Small to Medium Scale Instances} This section compares the performance of the heuristic solution with the exact solution obtained using Gurobi, focusing on both solution quality and computational efficiency. Given the computational complexity of solving the exact mathematical model, experiments are conducted on small to medium-scale instances ranging from $5$ to $35$ nodes, in increments of $5$ nodes. This range ensures that Gurobi can solve the model to optimality within a reasonable time frame. Table \ref{Small_Medium_Instances_Obj} presents the results, where the second and third columns report the average objective value $\mathcal{Z}_{\mu}^{e}$ and the standard deviation $\mathcal{Z}_{\sigma}^{e}$ of the optimal exact solution. Similarly, the fourth and fifth columns provide the average objective value $\mathcal{Z}_{\mu}^{h}$ and the standard deviation $\mathcal{Z}_{\sigma}^{h}$ for the heuristic solution. To quantify the percentage difference between the exact and heuristic solutions, a performance indicator referred to as \textit{Gap} $\Delta_{\mathcal{Z}}$ is computed using Equation (\ref{Gap}), where $\mathcal{Z}_{h}$ and $\mathcal{Z}_{e}$ denote the objective value of exact and heuristic solutions, respectively.
\begin{equation}
    Gap \; (\Delta_{\mathcal{Z}}) = \frac{\mathcal{Z}_{h}-\mathcal{Z}_{e}}{\mathcal{Z}_{e}} \times 100
    \label{Gap}
\end{equation}
The results, presented in Table \ref{Small_Medium_Instances_Obj} indicate that the proposed heuristic algorithm achieves near-optimal solutions, with an average gap of $\approx$$22\%$ compared to the exact solutions obtained using Gurobi. In terms of computation time, Table \ref{Small_Medium_Instances_Time} shows that as the number of customers increases, Gurobi requires significantly more time to compute optimal solutions. In contrast, the heuristic algorithm completes computations substantially faster while maintaining an acceptable solution quality. On average, the exact solution takes $1549.24$ seconds (s), whereas the heuristic solution requires only $0.043$s. These findings demonstrate the efficiency of the proposed approach for small to medium-scale instances.

\begin{figure}
\centering
\begin{minipage}{0.49\textwidth}
    \centering
    \includegraphics[width=\linewidth]{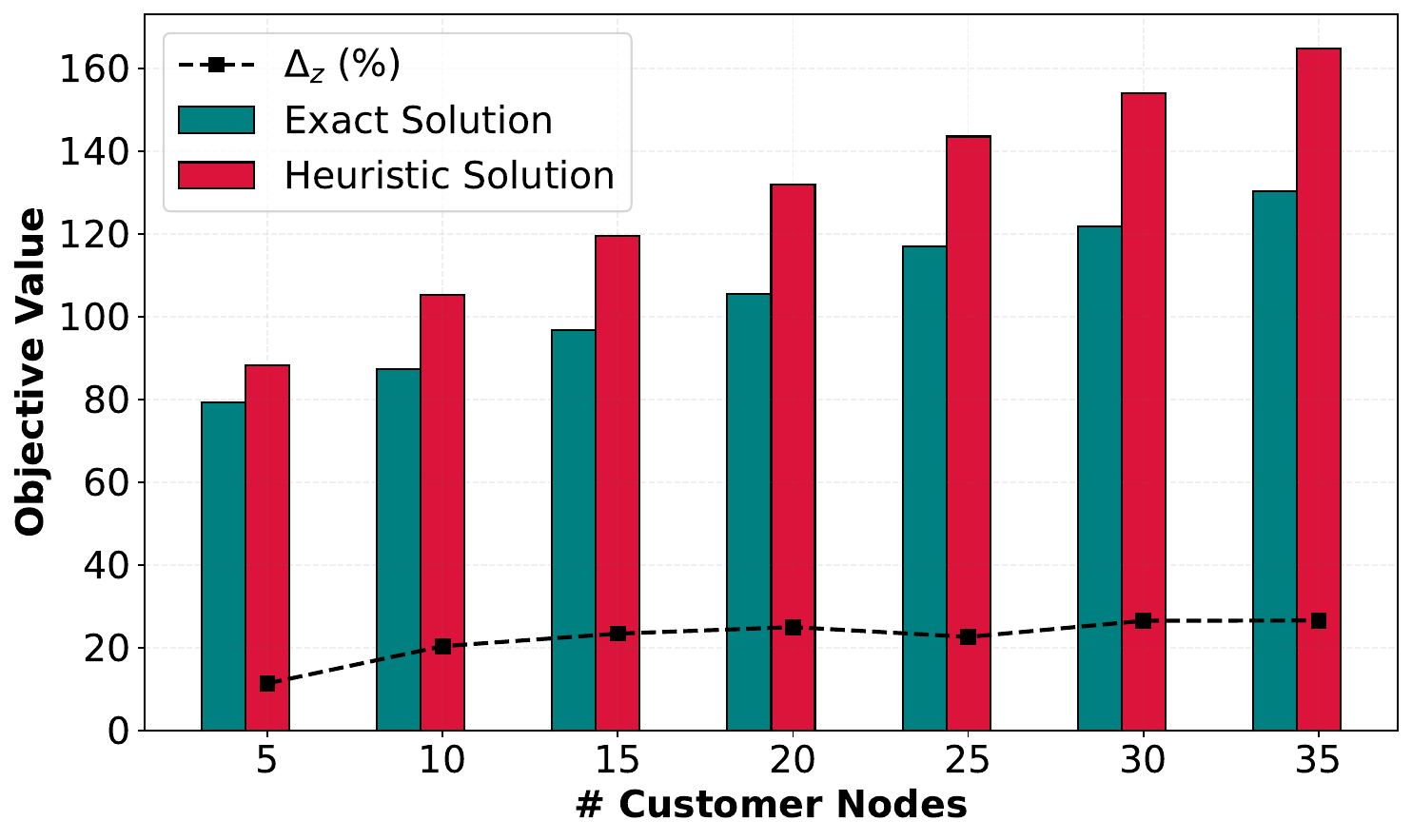}
    \caption{Objective values: exact vs heuristic (small/medium instances)}
    \label{ObjVal_small-medium_Instances}
\end{minipage}%
\hfill
\begin{minipage}{0.49\textwidth}
    \centering
    \includegraphics[width=\linewidth]{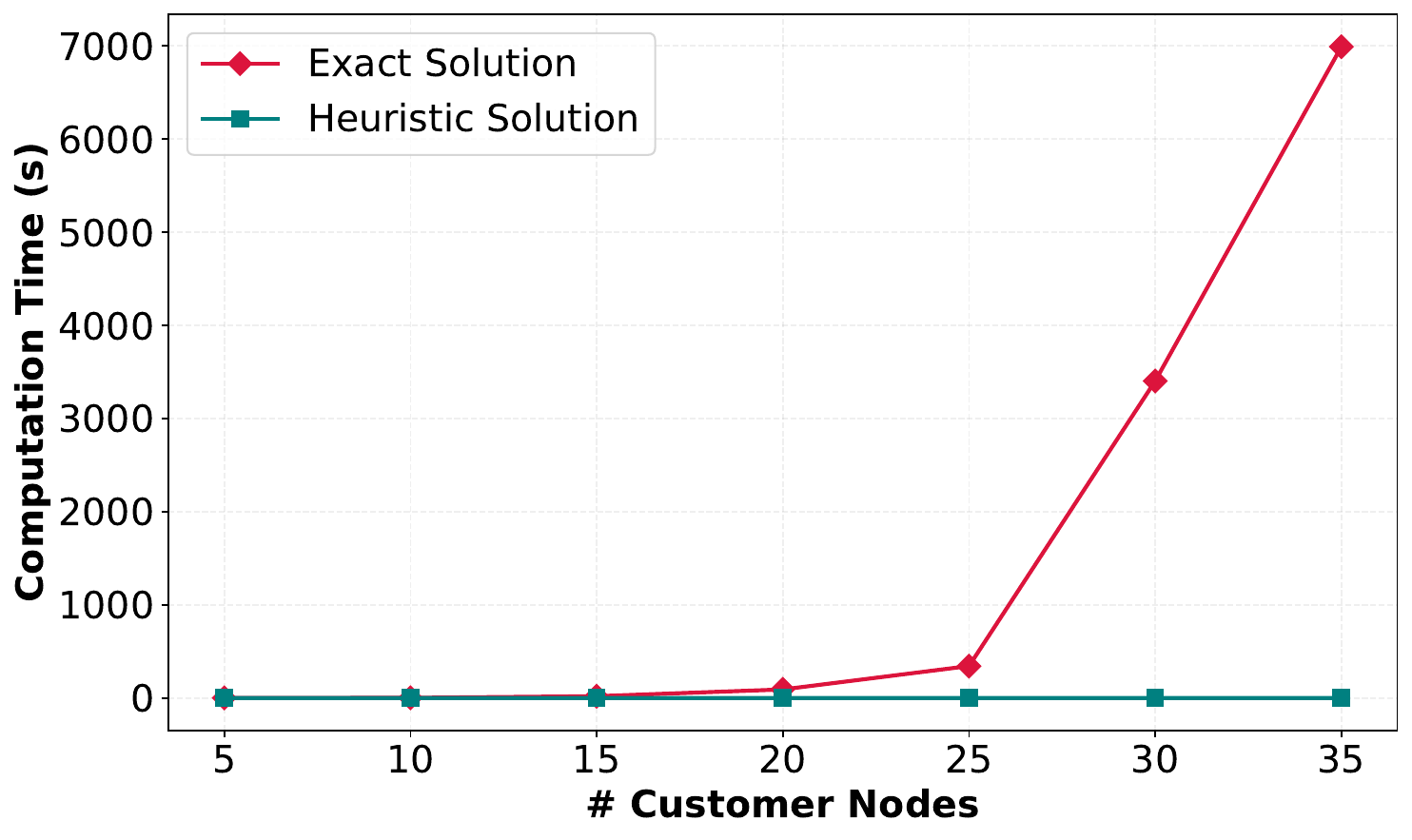}
    \caption{Computation times: exact vs heuristic (small/medium instances)}
    \label{ComputationTime_small-medium_Instances}
\end{minipage}
\end{figure}

Figures \ref{ObjVal_small-medium_Instances} and \ref{ComputationTime_small-medium_Instances} present a graphical comparison of objective values and computation times for exact and heuristic solutions across small and medium-scale instances.

\begin{table}[h]
\centering
\caption{Comparison of objective values $(\mathcal{Z})$ for small and medium scale instances between exact and heuristic solutions}
\begin{tabular}{c>{\centering\arraybackslash}p{1cm}>{\centering\arraybackslash}p{1cm}>{\centering\arraybackslash}p{1.2cm}>{\centering\arraybackslash}p{1cm}>{\centering\arraybackslash}p{0.7cm}}
\hline 
\multirow{2}{*}{{Scenario}} & \multicolumn{2}{c}{{Exact Solution }$(\mathcal{Z}_{e})$} & \multicolumn{2}{c}{{Heuristic Solution }$(\mathcal{Z}_{h})$} & \\ \cline{2-5} & $\mathcal{Z}_{\mu}^{e}$ & $\mathcal{Z}_{\sigma}^{e}$ & $\mathcal{Z}_{\mu}^{h}$ & $\mathcal{Z}_{\sigma}^{h}$ & {$\Delta_{\mathcal{Z}}\%$} \\ \hline
                     $S$$-$$5$  & \textbf{79.31} & 8.85 & 88.34 & 9.35 & 11.3 \\ 
                        $S$$-$$10$    & \textbf{87.40} & 8.18 & 105.21 & 10.27 & 20.3  \\ 
                        $S$$-$$15$    & \textbf{96.87 }& 8.56 & 119.55 & 9.53 & 23.4\\ 
                        $S$$-$$20$   & \textbf{105.51} & 6.82 &131.88 & 12.42 & 24.9\\ 
                        $M$$-$$25$   & \textbf{117.07} & 7.93 & 143.55 & 9.49& 22.6 \\ 
                         $M$$-$$30$   & \textbf{121.77 }& 5.11 & 154.06 & 10.10 & 26.5\\ 
                         $M$$-$$35$   & \textbf{130.23} & 6.52 & 164.85 & 9.95 & 26.5\\ \hline
                       
\end{tabular}
\label{Small_Medium_Instances_Obj}
\end{table}
\begin{table}[h]
\centering 
\caption{Comparison of computation time $(T)$ for small and medium scale instances between exact and heuristic solutions}
\begin{tabular}{c>{\centering\arraybackslash}p{1.2cm}>{\centering\arraybackslash}p{1.2cm}>{\centering\arraybackslash}p{1.2cm}>{\centering\arraybackslash}p{1.2cm}}
\hline 
\multirow{2}{*}{{Scenario}} & \multicolumn{2}{c}{{Exact Solution }$(T_{e})$} & \multicolumn{2}{c}{{Heuristic Solution }$(T_{h})$}  \\ \cline{2-5}  & $T_{\mu}^{e}$ & $T_{\sigma}^{e}$ & $T_{\mu}^{h}$ & $T_{\sigma}^{h}$ \\ \hline 
                        $S$$-$$5$   & 0.03 & 0.01 & \textbf{0} & 0 \\ 
                       $S$$-$$10$   & 1.64 & 1.23 & \textbf{0} & 0 \\ 
                        $S$$-$$15$   &19.08 & 11.70 & \textbf{0.01} & 0  \\ 
                        $S$$-$$20$   & 91.37 & 89.18 & \textbf{0.02} & 0.01 \\ 
                        $M$$-$$25$   & 342.28 & 195.3  & \textbf{0.05} & 0.02  \\ 
                       $M$$-$$30$   & 3402.62 & 1225.63 &  \textbf{0.08 }& 0.02   \\ 
                       $M$$-$$35$   & 6990.63 & 1500.32 &  \textbf{0.14}  & 0.08 \\ \hline
                        
\end{tabular}
\label{Small_Medium_Instances_Time}
\end{table}
\subsubsection{Experiment with Large-Scale Instances}
\begin{figure}[h]
    \centering
    \includegraphics[width=0.96\linewidth]{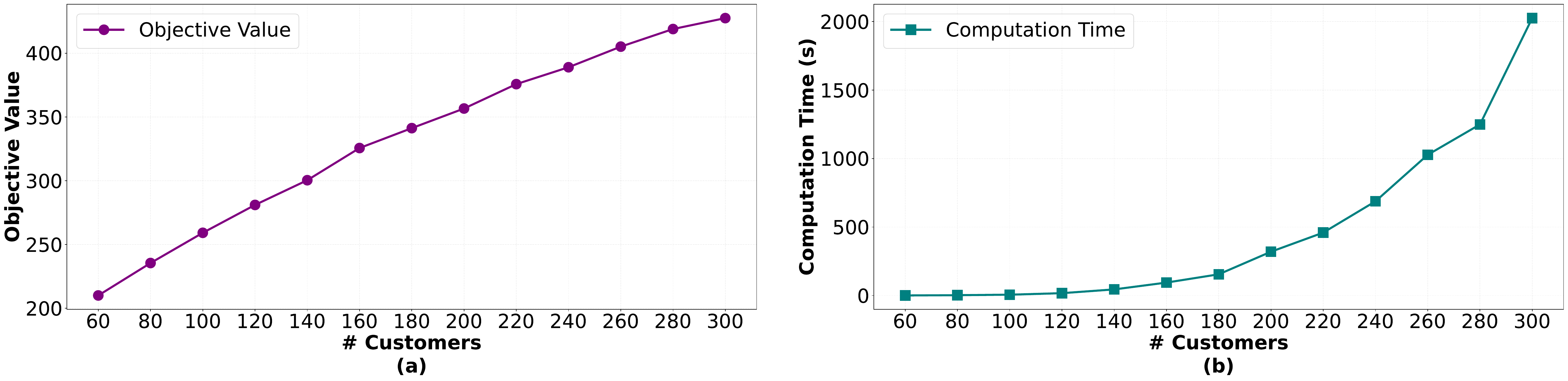}
    \caption{Graphical representation of the objective values and computation time of the heuristic solution for large-scale instances}
    \label{ObjVal_Time_large_Instances}
\end{figure}

This section evaluates the performance of the heuristic algorithm in solving large-scale instances. Experiments are conducted using the same parameter settings on instance sizes ranging from $60$ to $300$ nodes, in increments of $20$ nodes. For context, last-mile delivery trucks operated by companies such as FedEx and UPS typically manage approximately 100 stops per day \cite{yanpirat2023sustainable}.Tables \ref{Large_Instances_Obj} and \ref{Large_Instances_Time} summarize the average objective values and computation times for the experiments. As expected, the computation time of the heuristic increases with the size of the problem. The average computational times for instances with $100$, $200$, and $300$ nodes are $6.30$s, $319.95$s, and $2025.20$s, respectively, demonstrating that the heuristic remains computationally efficient even for larger problem scales. These results highlight the effectiveness of the proposed solution approach in handling large-scale instances, providing near-optimal solutions within reasonable time frames. Figure \ref{ObjVal_Time_large_Instances}(a) and Figure \ref{ObjVal_Time_large_Instances}(b) provide a graphical representation of objective values and computation times, respectively. 

\begin{table}[h]
\centering
\begin{minipage}{0.48\textwidth}
    \centering
    \caption{Objective values $(\mathcal{Z})$ of the heuristic solution for large-scale instances}
    \resizebox{\textwidth}{!}{%
    \begin{tabular}{>{\centering\arraybackslash}p{2cm}
                    >{\centering\arraybackslash}p{2cm}
                    >{\centering\arraybackslash}p{2cm}}
    \hline 
    \multirow{2}{*}{{Scenario}} & \multicolumn{2}{c}{{Heuristic Solution} $(\mathcal{Z}_{h})$}  \\ 
    \cline{2-3}  & $\mathcal{Z}_{\mu}^{h}$ & $\mathcal{Z}_{\sigma}^{h}$ \\ \hline 
    $L$-$60$    & 210.01 & 12.25\\ 
    $L$-$80$    & 235.44 & 12.02 \\ 
    $L$-$100$   & 259.14 & 9.93 \\ 
    $L$-$120$   & 281.01 & 8.04 \\ 
    $L$-$140$   & 300.40 & 13.16 \\ 
    $L$-$160$   & 325.63 & 13.03 \\ 
    $L$-$180$   & 341.24 & 10.05 \\ 
    $L$-$200$   & 356.63 & 8.19 \\ 
    $L$-$220$   & 375.75 & 9.93 \\ 
    $L$-$240$   & 389.00 & 10.60 \\ 
    $L$-$260$   & 405.16 & 8.28 \\ 
    $L$-$280$   & 418.96 & 12.24 \\ 
    $L$-$300$   & 427.53 & 10.87\\ 
    \hline
    \end{tabular}
    }
    \label{Large_Instances_Obj}
\end{minipage}%
\hfill
\begin{minipage}{0.48\textwidth}
    \centering
    \caption{Computation time $(T)$ of the heuristic solution for large-scale instances}
    \resizebox{\textwidth}{!}{%
    \begin{tabular}{>{\centering\arraybackslash}p{2cm}
                    >{\centering\arraybackslash}p{2cm}
                    >{\centering\arraybackslash}p{2cm}}
    \hline 
    \multirow{2}{*}{{Scenario}} & \multicolumn{2}{c}{{Heuristic Solution} $(T_{h})$}  \\ 
    \cline{2-3}  & $T_{\mu}^{h}$ & $T_{\sigma}^{h}$ \\ \hline 
    $L$-$60$    & 0.73 & 0.30\\ 
    $L$-$80$    & 2.68 & 0.94 \\ 
    $L$-$100$   & 6.30 & 2.10 \\ 
    $L$-$120$   & 17.39 & 6.89 \\ 
    $L$-$140$   & 44.95 & 8.60 \\ 
    $L$-$160$   & 94.70 & 19.15\\ 
    $L$-$180$   & 155.10 & 46.90\\ 
    $L$-$200$   & 319.95 & 170.6 \\ 
    $L$-$220$   & 459.39 & 129.68\\ 
    $L$-$240$   & 687.92 & 154.04\\ 
    $L$-$260$   & 1027.48 & 175.52\\ 
    $L$-$280$   & 1248.99 & 185.23\\ 
    $L$-$300$   & 2025.20 & 420.58\\ 
    \hline
    \end{tabular}
    }
    \label{Large_Instances_Time}
\end{minipage}
\end{table}

\subsection{Experiments with Varying Collaborative Modes}
This subsection analyzes the operational cost and makespan of different fleet compositions relative to truck-only (TO) deliveries, assuming a classical vehicle routing problem context. Three collaborative modes are evaluated: i) truck and drone (TD), ii) truck and robot (TR) and iii) the entire fleet comprised of truck, drone, and robot (EF). The experiments are conducted across instances ranging from $20$ to $300$ nodes.

\begin{figure}[h]
    \centering
    \includegraphics[width=0.99\linewidth]{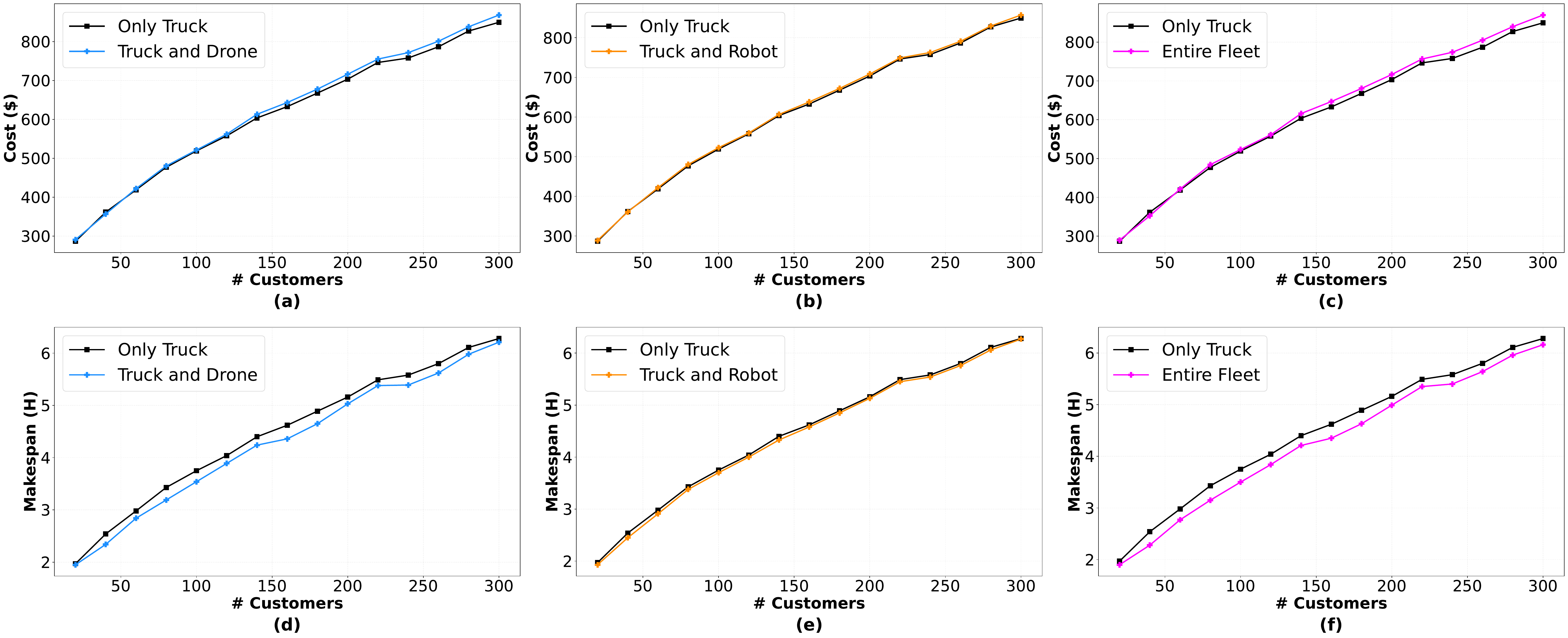}
    \caption{Impact of varying collaborative modes on operational cost and makespan compared to the truck-only mode}
    \label{Multimode_delivery_Comparison}
\end{figure}

Figures \ref{Multimode_delivery_Comparison}(a)–\ref{Multimode_delivery_Comparison}(c) compare the operational costs of collaborative modes against TO.  Surprisingly, all collaborative modes exhibit slight cost increases: $1.18\%$, $0.56\%$, and $1.30\%$, for TD, TR, and EF, respectively, compared to the TO scenario. These marginally higher costs are attributed to the added complexity of coordinating and integrating auxiliary vehicles into the delivery process. However, in Figure \ref{Multimode_delivery_Comparison}(d)–\ref{Multimode_delivery_Comparison}(f), the makespan of all collaborative modes consistently outperforms the TO case, emphasizing the efficiency gains achieved through multi-modal delivery systems. The TD mode reduces the makespan by an average of $3.89\%$, demonstrating drones’ advantage in expedited deliveries. Similarly, the TR mode achieves an average improvement of $1.24\%$, showing that robots, while effective, may not provide as substantial a time-saving benefit as drones. The EF configuration yields the highest improvement, reducing makespan by $4.82\%$, highlighting the synergistic benefits of a fully collaborative fleet. These findings emphasize a trade-off between cost and efficiency in multi-platform last-mile delivery. While cost reductions are minimal, the substantial makespan improvements underscore the potential value of collaborative fleets in time-sensitive or high-demand delivery scenarios.

\subsection{Experiments with Single-Visit and Multiple-Visits of Drones and Robots}
\begin{figure}
    \centering
    \includegraphics[width=0.7\linewidth]{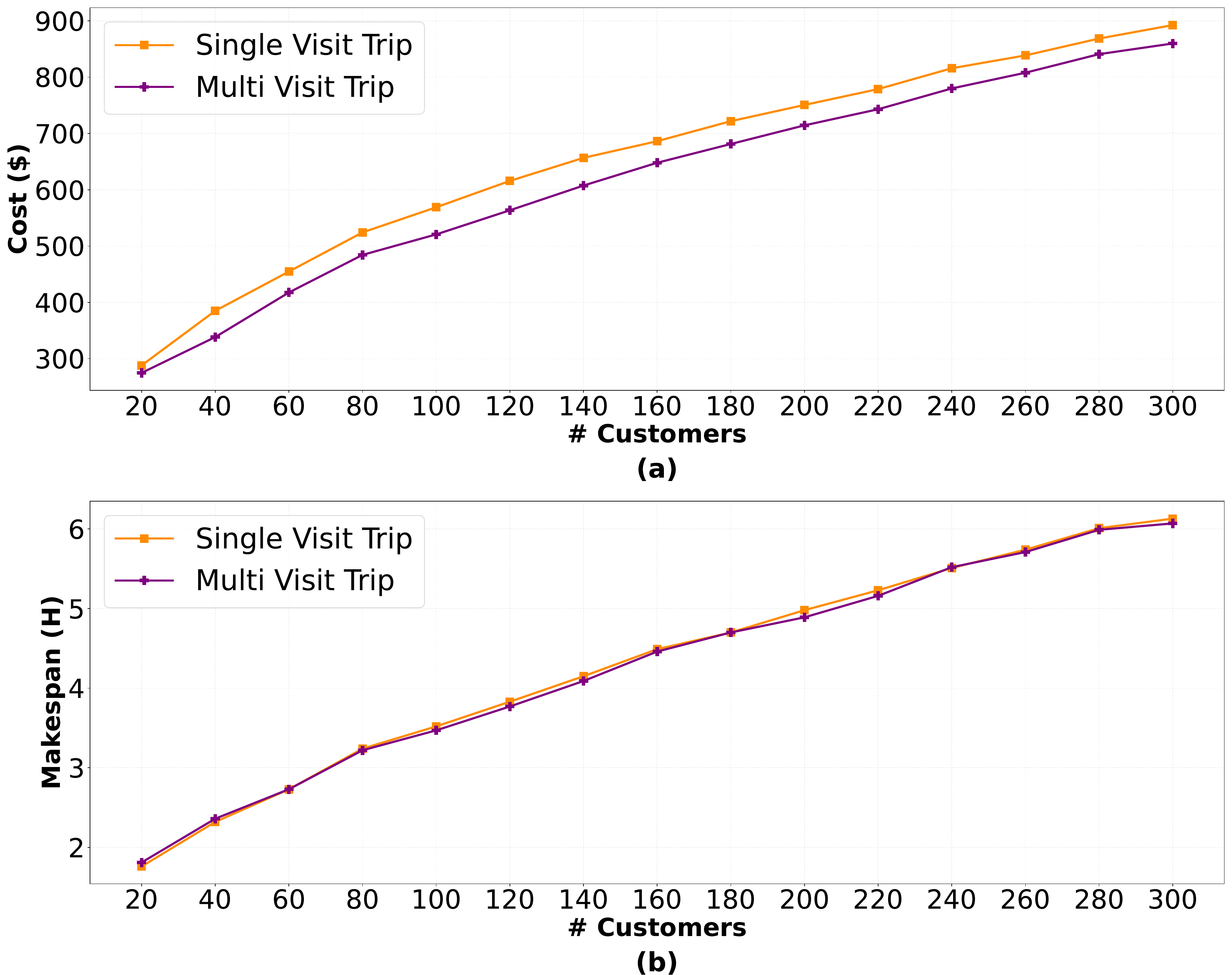}
    \caption{Impact of single-visit and multi-visit of drones and robots on operational cost and makespan}
    \label{Time_Cost_Single_Multi-Vist}
\end{figure}

This experiment assesses the impact of allowing drones and robots to serve multiple customers per trip (multi-visit mode) compared to the single-visit scenario, where each trip is restricted to a single customer. To ensure a fair comparison, customer node locations are identically generated using a fixed random seed, maintaining consistent geographic distribution across all instances. Experiments are conducted on instances ranging from $20$ to $300$ nodes, incremented in intervals of $20$ nodes. The average results, summarized in Table \ref{tbl_single_multi_visit}, compare operational costs $\mathcal{C}_{s}, \mathcal{C}_{m}$ and makespan $\Gamma_{s}, \Gamma_{m}$ for the single-visit and multi-visit scenarios, respectively. The cost gap $\Delta_{\mathcal{C}}$ between the two cases is calculated using Equation (\ref{Gap2}). 
\begin{equation}
    Gap \; (\Delta_{\mathcal{C}}) = \frac{\mathcal{C}_{s}-\mathcal{C}_{m}}{\mathcal{C}_{m}} \times 100
    \label{Gap2}
\end{equation}
Results demonstrate that enabling multi-visit functionality provides significant advantages, particularly in terms of cost reduction. On average, the single-visit scenario incurs a $6.64\%$ higher operational costs than the multi-visit approach. This cost reduction in the multi-visit mode is attributed to the ability of drones and robots to serve multiple customers in a single trip, minimizing the total distance traveled by the truck. Interestingly, despite the cost savings, there is a negligible difference in makespan between the two scenarios. The comparative results for operational costs and makespan are visually presented in Figures \ref{Time_Cost_Single_Multi-Vist}(a) and Figure \ref{Time_Cost_Single_Multi-Vist}(b).

\begin{table}
\centering
\caption{Comparison of operational cost and makespan between single-visit and multi-visits}
\begin{tabular}{c>{\centering\arraybackslash}p{1cm}>{\centering\arraybackslash}p{1cm}>{\centering\arraybackslash}p{1cm}>{\centering\arraybackslash}p{1cm}>{\centering\arraybackslash}p{1cm}}
\hline 
\multirow{2}{*}{{Scenario}} & \multicolumn{3}{c}{{Operational Cost} $(\mathcal{C})$} & \multicolumn{2}{c}{{Makespan} $(\Gamma)$ }  \\ \cline{2-6}  & $\mathcal{C}_{s}$ & $\mathcal{C}_{m}$ & {$\Delta_{\mathcal{C}}\%$} &  $\Gamma_{s}$ & $\Gamma_{m}$ \\ \hline 
                         $S$$-$$20$    & 288.17  & \textbf{274.85}  &   4.85 &  \textbf{1.76 } & 1.81    \\ 
                         $L$$-$$40$    & 385.15  & \textbf{338.50}  &  13.78   & \textbf{2.32}   & 2.36   \\ 
                         $L$$-$$60$    & 454.85  & \textbf{417.68}  & 8.90   & 2.73   & 2.73   \\ 
                         $L$$-$$80$    & 524.41  & \textbf{484.43 } &  8.25  & 3.24   & \textbf{3.22}    \\ 
                         $L$$-$$100$   & 569.01  & \textbf{520.82}  &  9.25  & 3.52   & \textbf{3.47 }  \\ 
                         $l$$-$$120$   & 615.83  & \textbf{563.65 } &  9.26  & 3.83   &  \textbf{3.77}  \\ 
                         $L$$-$$140$   & 656.81  & \textbf{607.56 } &  8.11  & 4.15   & \textbf{4.09 }  \\ 
                         $L$$-$$160$   & 686.34  & \textbf{648.05 } & 5.91   & 4.49   & \textbf{4.46}   \\ 
                         $L$$-$$180$   & 721.92  & \textbf{681.37}  &  5.95  & 4.70   & 4.70    \\ 
                         $L$$-$$200$   & 750.73  & \textbf{714.47 } &  5.08  & 4.98   & \textbf{4.89 }  \\ 
                         $L$$-$$220$   & 778.87  &  \textbf{743.06} &  4.82   & 5.23   & \textbf{5.16 }  \\ 
                         $L$$-$$240$   & 815.87  &  \textbf{780.00} &   4.60   & \textbf{5.51 }  & 5.52   \\ 
                         $L$$-$$260$   & 838.71  &  \textbf{807.94} &  3.81     & 5.74   & \textbf{5.71 }  \\ 
                         $L$$-$$280$   & 868.67  &  \textbf{840.93} &   3.30    & 6.01   &  \textbf{5.99 } \\ 
                         $L$$-$$300$   & 892.54  &  \textbf{859.86 }&   3.80    & 6.13   & \textbf{6.07}   \\ \hline
                       
\end{tabular}
\label{tbl_single_multi_visit}
\end{table}

\subsection{Experiments with Single and Multiple Trips of Drones and Robots}
\begin{figure}[h]
    \centering
    \includegraphics[width=0.7\linewidth]{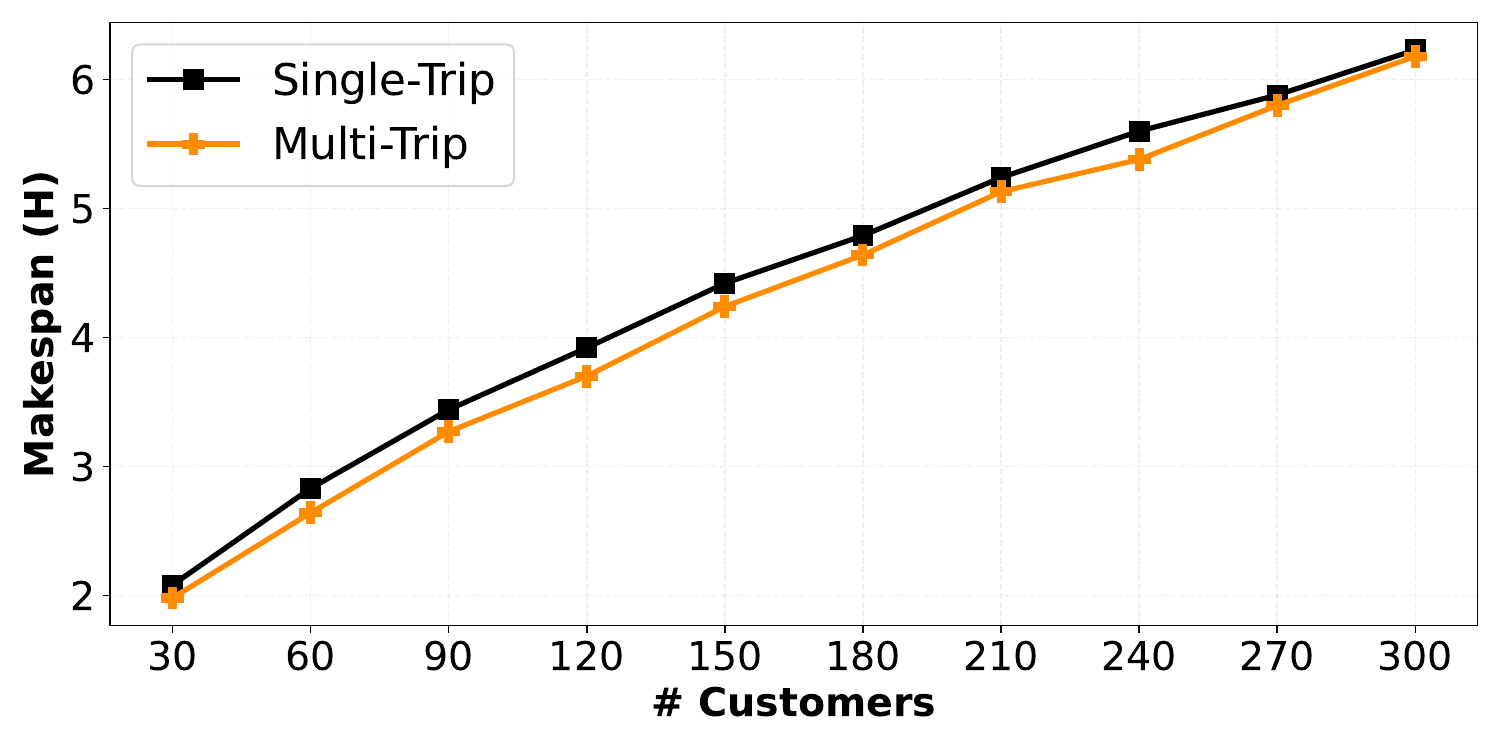}
    \caption{Impact of single-trip and multi-trips of drones and robots on makespan}
    \label{single_vs_multi_trip}
\end{figure}

This experiment evaluates the impact of allowing drones and robots to make multiple trips (multi-trip mode) versus the single-trip scenario, where each drone and robot is restricted to exactly one trip. The experiments are conducted on instances ranging from $30$ to $300$ nodes, incremented in intervals of $30$ nodes. The results presented in Figure \ref{single_vs_multi_trip} show that multi-trip functionality significantly reduces makespan. On average, the multi-trip scenario achieves a $4.0\%$ reduction in makespan compared to the single-trip mode.
\subsection{Analysis of the Impact of the En-route Charging Feature}
\begin{figure}[h]
    \centering
    \includegraphics[width=0.7\linewidth]{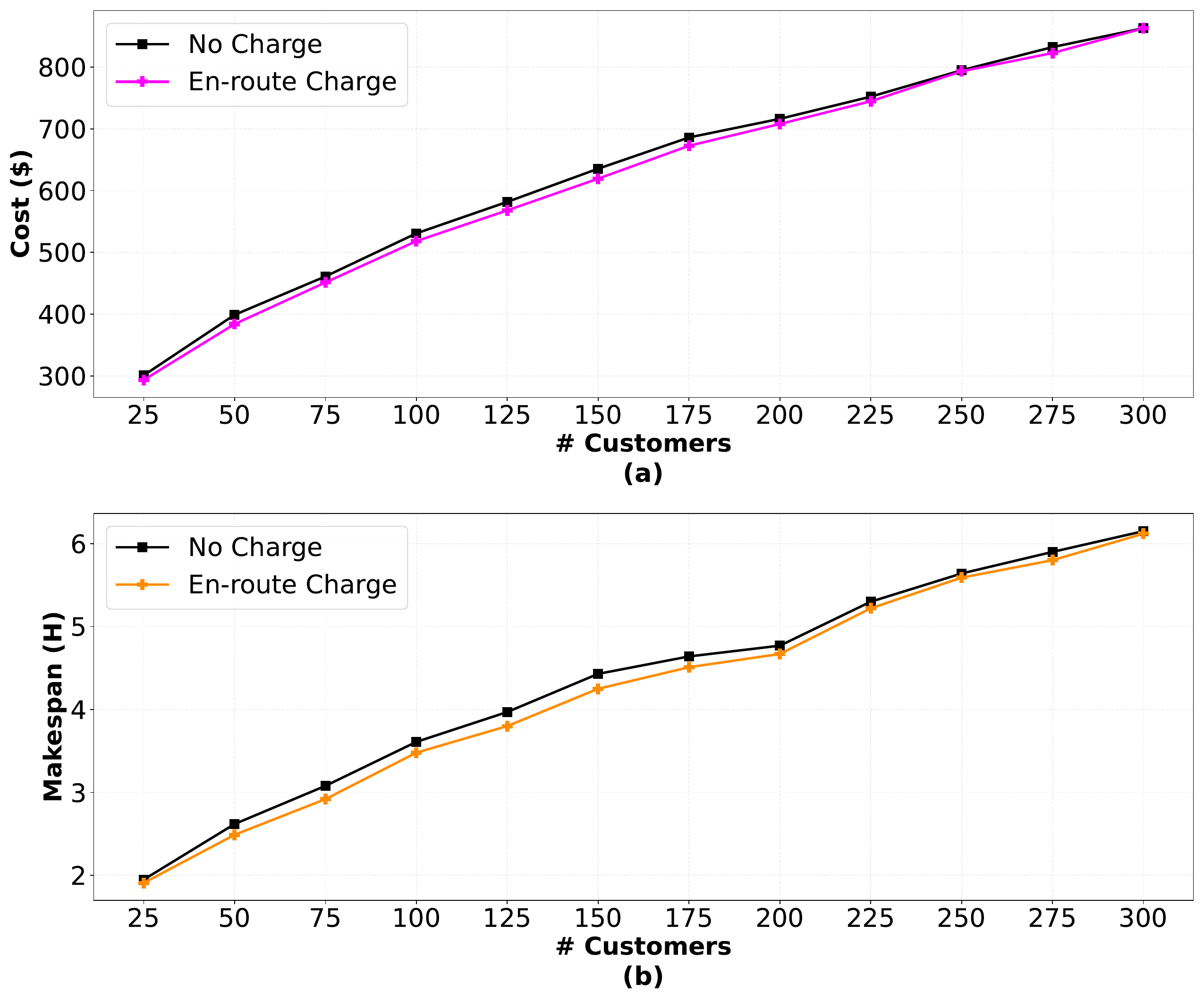}
    \caption{Comparison of en-route charging and no-charging modes in terms of cost and makespan}
    \label{enRoute_No_charge}
\end{figure}

This section analyzes the impact of en-route charging and no-charge modes of vehicles. The en-route charging mode is the focus of this study here. The no-charge mode allows the drone and robot to be recharged only at the depot. A comparative experiment is conducted across varying instance scales to evaluate the two modes' effectiveness. The results, presented in Figure \ref{enRoute_No_charge}, indicate that the en-route charging mode reduces the average operational cost by $\approx$$1.80\%$ and the makespan by around $2.91\%$ compared to the no-charge mode. The results align with findings from Yu et al. \cite{yu2022electric}, who studied en-route charging exclusively for robots. Their results similarly demonstrated a cost reduction of $\approx$$0.6\%$ when switching from a static-charge mode (recharging only at designated stations) to an en-route charging mode. The efficiency gains of en-route charging depend on factors such as charging rates and battery capacities of different vehicles. While the $\approx$$1.80\%$ cost reduction represents notable savings in last-mile delivery, the logistics companies should consider adopting en-route charging only if the fixed cost of implementing this technology remains feasible.

\subsection{Flexible Docking}
\begin{figure}[h]
    \centering
    \includegraphics[width=0.99\linewidth]{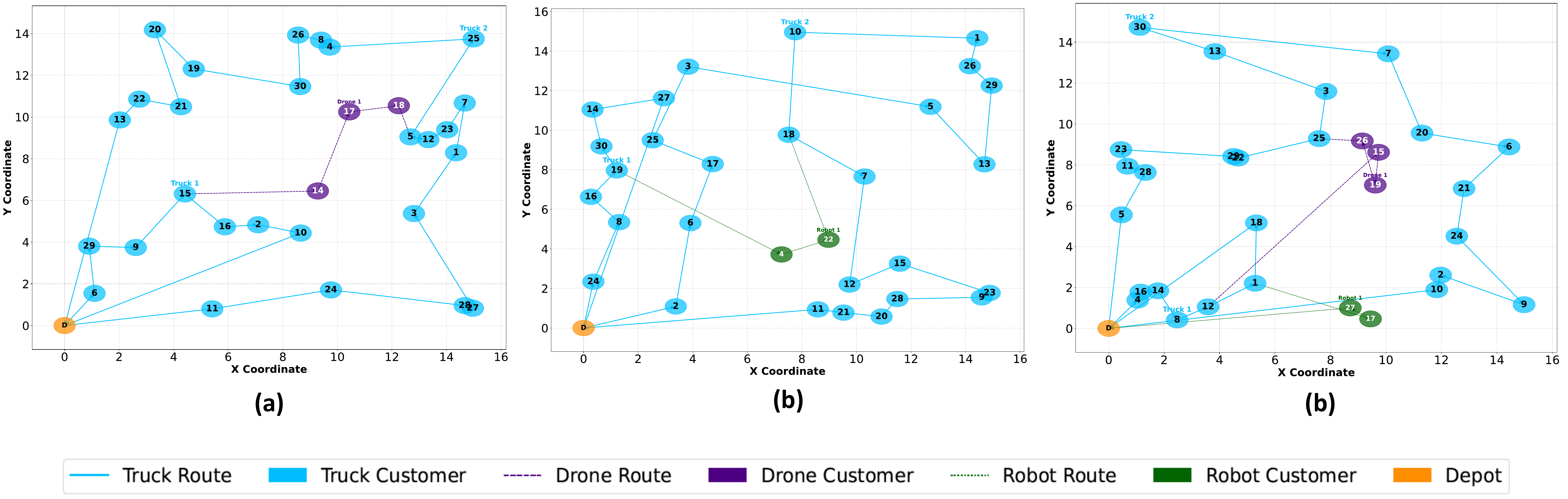}
    \caption{Visualization of flexible docking scenarios}
    \label{flexible_docking}
\end{figure}

This section illustrates the flexible docking concept, which allows drones and robots to return to any available nearby truck upon completing their missions, rather than being restricted to a fixed truck assignment. The experiment is conducted with a fleet of two trucks, one drone, and one robot, serving $30$ customer nodes. Figure \ref{flexible_docking}(a) demonstrates a drone deployment scenario. Drone $1$ is launched from Node $15$ on Truck $1$’s route, serves customers $\left< 14, 17, 18 \right>$, and subsequently returns to Node $5$ on Truck $2$’s route. This highlights the operational flexibility afforded by the system. Similarly, Figure \ref{flexible_docking}(b) depicts a robot deployment. The robot is launched from Node $19$ on Truck $1$’s route, serves customers $\left< 4, 24 \right>$, and returns to Node $18$ on Truck $2$’s route, further showcasing the adaptability of the flexible docking approach. In a combined deployment scenario, shown in Figure \ref{flexible_docking}(c), the drone launches from Node $12$ on Truck $1$'s route, serves customers $\left< 15, 19, 26 \right>$, and returns to Node $25$ on Truck $2$’s route. Concurrently, the robot is deployed from Node $1$ on Truck $1$’s route, serves customers $\left< 17, 27 \right>$, and returns directly to the depot. These visualizations underscore the effectiveness of the flexible docking concept, which enables more efficient utilization of fleet and enhances the overall system performance through dynamic deployment.
\subsection{Sensitivity Analysis}
This subsection presents a comprehensive sensitivity analysis to assess the impact of key parameters on system performance and derive some managerial insights. The analysis is conducted on both small- and large-scale instances to capture diverse operational scenarios. The key parameters evaluated include speed, payload, distance, and the impacts of varying the number of drones.

\subsubsection{Impact of Varying Number of Drones Carried by Truck}
\begin{figure}
    \centering
    \includegraphics[width=0.7\linewidth]{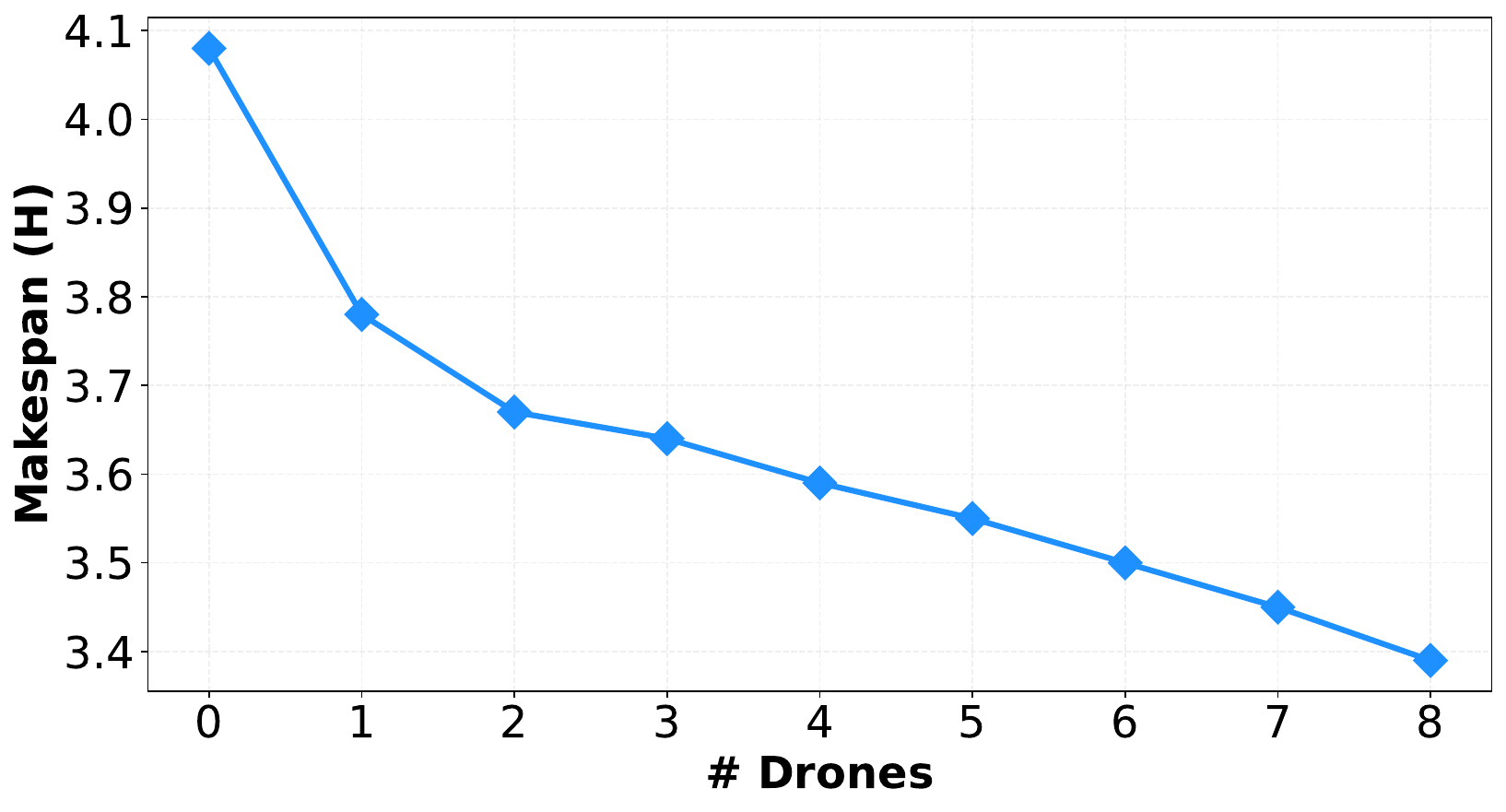}
    \caption{The makespan results with varying numbers of drones}
    \label{Number_of_Drones}
\end{figure}

This experiment evaluates how changing the number of drones deployed per truck affects system performance. A large-scale instance with $120$ customers is used, and the number of drones $\left( \left| \mathcal{D} \right| = 0,1,2, ...,8 \right)$ is systematically varied. The average of the results are graphically illustrated in Figure \ref{Number_of_Drones}. Figure \ref{Number_of_Drones} highlights the reduction in makespan compared to the truck-only case as $ \left| \mathcal{D} \right|$ increases. Without drones $\mathcal{D}=0$, the makespan is $4.08$h, but it decreases to $3.39$h with eight drones $\mathcal{D}=8$, representing a substantial reduction of $\approx$$16.91\%$ in delivery time. The results indicate that the most noticeable reduction occurs when adding the first drone, which decreases the makespan by $\approx$$7.35\%$. However, subsequent drone additions yield diminishing returns, with the reduction from seven to eight drones being $\approx$$1.74\%$. While assigning more drones reduces delivery time, it also escalates operational costs. Therefore, logistics planners must strike a balance between deploying additional drones and meeting customer demand to achieve an efficient and cost-effective delivery system.

\subsubsection{Impact of Fleet Speed}
\begin{figure*}
    \centering
    \includegraphics[width=0.99\linewidth]{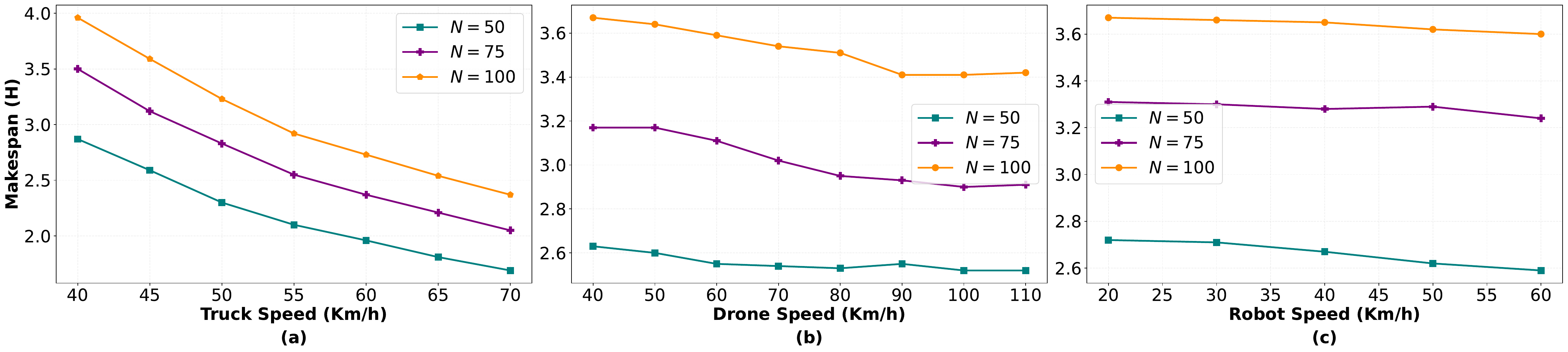}
    \caption{Impact of fleet speed on makespan}
    \label{Fleet_Speed_Comparison}
\end{figure*}

Speed plays a critical role in determining the efficiency of parcel delivery in last-mile logistics. This experiment analyzes the impact of varying the speeds of the truck, drone, and robot on the makespan required to complete all deliveries, while keeping all other conditions constant. To capture different delivery scenarios, the analysis is conducted across three customer instances: $\mathcal{N}=50$, $75$ and $100$. Each fleet operates within a specific speed range. The experimental results are visually summarized in Figure \ref{Fleet_Speed_Comparison}.

In Figure \ref{Fleet_Speed_Comparison}(a), the results show that increasing truck speed significantly reduces makespan across all customer scenarios. For instance, increasing the truck's speed from $40$ km/h to $70$ km/h results in a makespan reduction of $\approx$$40.8\%$, depending on the number of customers. This substantial improvement is due to the truck's central role in synchronizing auxiliary vehicle operations. A faster truck minimizes waiting times for drones and robots, enabling them to dispatch and rendezvous more efficiently. For $\mathcal{N}=50$, the makespan drops from $2.87$h to $1.69$h, reflecting a reduction of $\approx$$41.1\%$. Similarly, for $\mathcal{N}=100$, the makespan decreases from $3.96$h to $2.37$h, demonstrating a $\approx$$40.1\%$ improvement.

Figure \ref{Fleet_Speed_Comparison}(b) illustrates that increasing drone speed from $40$ km/h to $110$ km/h reduces makespan by $\approx$$4.18\%–6.18\%$, with the most significant improvements occurring between $40–70$ km/h. For instance, with $\mathcal{N}=75$, the makespan drops from $3.17$h to $3.02$h, marking a $\approx$$4.3\%$ improvement. For $\mathcal{N}=100$, the reduction is more significant, with the makespan decreasing from $3.67$h to $3.54$h, a $\approx$$3.54\%$ improvement. However, beyond $90$ km/h, the benefits diminish due to payload and battery capacity constraints, which limit the drones' ability to serve more customers efficiently.

Figure \ref{Fleet_Speed_Comparison}(c) shows that increasing robot speed from $20$ km/h to $60$ km/h results in a modest makespan reduction of $\approx$$1.90\%-4.77\%$. This smaller impact is expected, as robots are typically tasked with short-distance deliveries. Even with a significant speed increase, the overall system performance improves only marginally, especially compared to the pronounced improvements gains from faster trucks or drones.
\subsubsection{Impact of Drone Payload Capacity}
Increasing the drone payload capacity improves operational efficiency by allowing drones to carry more parcels per trip, thereby reducing overall delivery time. To analyze the impact of payload capacity on makespan, experiments are conducted with payloads ranging from $5$ kg to $30$ kg in $5$ kg increments across three customer scenarios: $\mathcal{N}=50$, $\mathcal{N}=75$, and $\mathcal{N}=100$. The results, as presented in Figure \ref{Drone_Payload)_Distance}(a), indicate that a higher payload capacity consistently reduces the makespan for all scenarios. For instance, when $\mathcal{N}=50$, the makespan decreases from $\approx$$2.73$h at $5$ kg payload to $2.52$h at $20$ kg payload, reflecting a $\approx$$7.69\%$ reduction. Similarly, for $\mathcal{N}=75$, the makespan reduces from $3.19$h at $5$ kg payload to $3.03$h at $20$ kg payload, yielding a $\approx$$5.0\%$ reduction, while for $\mathcal{N}=100$, the makespan drops from $3.67$h at $5$ kg payload to $3.47$h at $20$ kg payload, indicating a $\approx$$5.5\%$ reduction. However, the reduction in makespan diminishes as the payload capacity increases, particularly beyond the range of $20$ kg range. However, the reduction in makespan diminishes beyond 20 kg, as energy and distance constraints begin to limit the drones' ability to serve additional customers effectively.

\subsubsection{Impact of Drone Flying Distance}
\begin{figure}[h]
    \centering
    \includegraphics[width=0.7\linewidth]{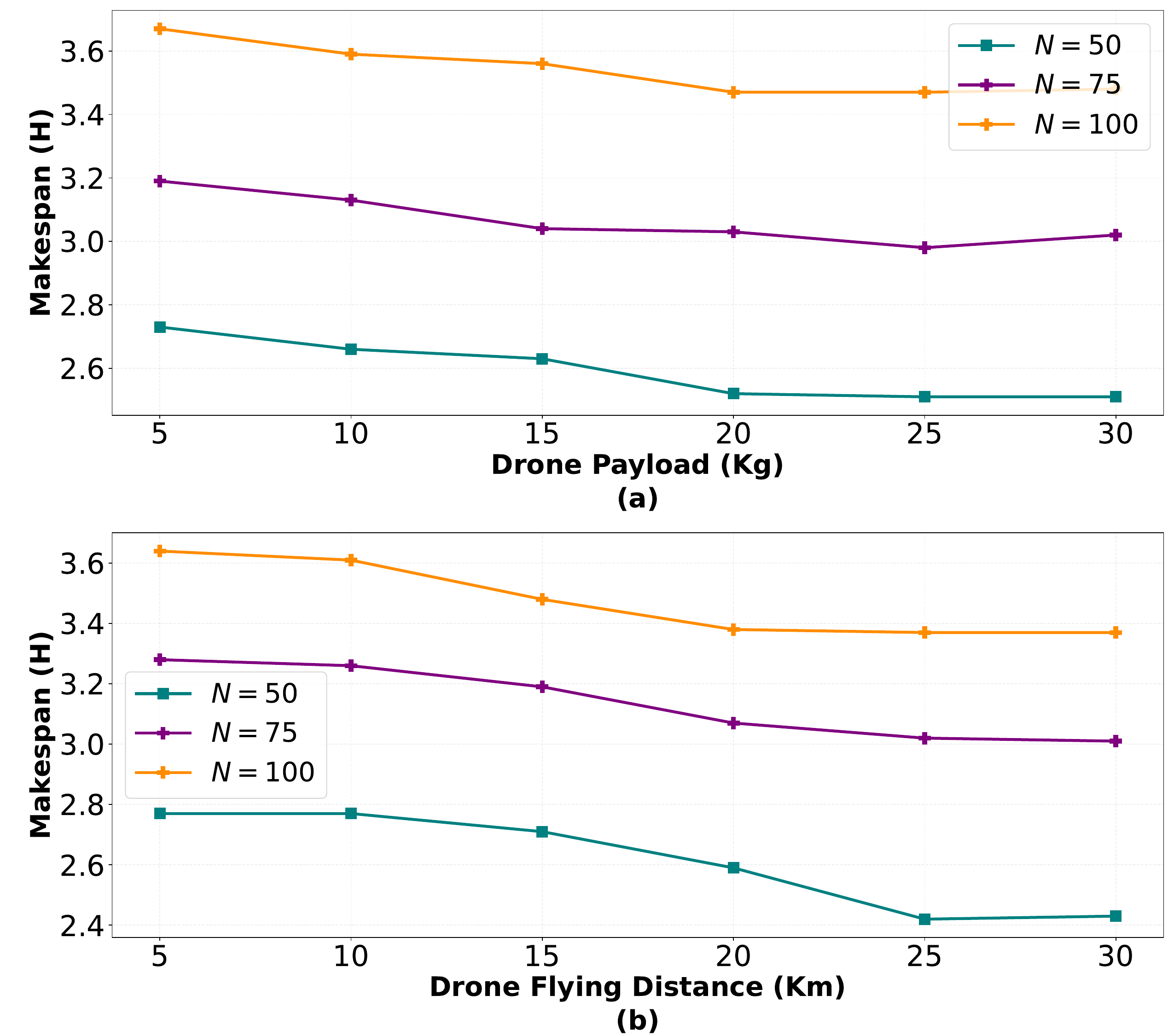}
    \caption{(a) Impact of varying drone payload capacity on makespan (b) Impact of varying drone distance on makespan}
    \label{Drone_Payload)_Distance}
\end{figure}

The drone flying distance significantly impacts customer allocation and overall delivery efficiency. With advancements in drone technology, extending the maximum travel range has become feasible \cite{wang2019vehicle}. This experiment investigates the impact of varying drone flying distance from $5$ km to $30$ km on the total makespan across different scenarios. The results presented in Figure \ref{Drone_Payload)_Distance}(b) indicate that increasing the maximum drone range leads to a consistent reduction in makespan across all scenarios. For instance, when $\mathcal{N}=50$, the makespan decreases from $2.77$h at $5$km to $2.59$h at $20$ km, representing a $\approx$$6.5\%$ reduction. Similarly, for $\mathcal{N}=75$, the makespan drops from $3.28$h at $5$ km to $3.07$h at $20$ km, yielding a $\approx$$6.4\%$ reduction, while for $\mathcal{N}=100$, it decreases by $\approx$$7.1\%$. However, beyond 20 km, the reduction in makespan becomes marginal, indicating diminishing returns. This suggests that factors such as payload limitations and energy constraints restrict the drones from fully utilizing their extended flying range, thereby limiting further operational benefits.

\section{Conclusion}
\label{conclusion}
This paper presents the collaborative multi-platform vehicle routing problem with multiple trucks, drones, and robots (VRP-DR), designed to enhance last-mile logistics efficiency. In this framework, trucks act as mobile platforms, enabling drones and robots to launch, retrieve, and perform en-route charging. The VRP-DR accounts for realistic constraints for payload, distance, and energy consumption for drones and robots. Each delivery begins with trucks departing from a central depot, carrying parcels along with drones and robots. Each customer is served exactly once by either, a truck, drone, or robot, and all vehicles must return to the depot after completing their deliveries. The model supports multi-trip operations, multi-visit services, flexible docking, and both cyclic and acyclic operations. The VRP-DR is formulated as MILP to minimize both operational costs and makespan. While the MILP can only solve small-scale instances optimally using the Gurobi, a scalable heuristic algorithm, FINDER, is developed to efficiently solve large-scale instances. Computational experiments conducted on small-scale instances ($5–35$ nodes) demonstrate that the heuristic solution provides near-optimal solutions with significantly reduced computation time compared to the exact MILP approach. For larger instances ($60–300$ nodes), the heuristic proves to be scalable, solving the problem within acceptable timeframes.

Numerical results reveal that collaborative delivery modes, which involve truck-drone, truck-robot, or the entire fleet, incur slightly higher costs compared to truck-only scenarios due to the need for enhanced coordination and integration of drones and robots. However, these collaborative modes consistently achieve substantial reductions in makespan, with the entire fleet configuration reducing makespan by $4.82\%$ compared to the truck-only case. Experiments comparing single-visit and multi-visit scenarios indicate that single-visit operations result in an average cost increase of $6.64\%$. Additionally, enabling en-route charging leads to a $1.80\%$ reduction in operational costs and a $2.91\%$ decrease in makespan, with even greater benefits expected in wider delivery areas. Flexible docking further enhances fleet utilization efficiency, as highlighted in visualizations. A sensitivity analysis demonstrates that fleet speed, payload capacity, and the number of drones per truck significantly influence performance, with moderate increases in these parameters leading to notable makespan reductions, though diminishing returns occur beyond certain thresholds.

For future research, several extensions are proposed to further enhance the applicability of the VRP-DR model. First, the model can be extended to simultaneously handle both pickup and delivery operations in collaborative systems. Second, incorporating heterogeneous fleets of trucks, drones, and robots, along with multiple depots, would broaden the operational scope and accommodate diverse vehicle capabilities and constraints. Third, a more detailed representation of launch and retrieval operations at discrete points along delivery routes could improve the accuracy of fleet coordination. Lastly, integrating dynamic factors such as weather conditions, fuel consumption, and customer time windows would enhance real-world applicability. 

\section*{CRediT authorship contribution statement}
\textbf{Sumbal Malik:} Writing – original draft, Conceptualization, Methodology, Visualization. \textbf{Majid Khonji:} Writing – review \& editing, Methodology, Conceptualization. \textbf{Khaled Elbassioni:} Writing – review \& editing, Methodology, Conceptualization. \textbf{Jorge Dias:} Writing – review \& editing.

\section*{Declaration of competing interest}
The authors declare that they have no known competing financial interests or personal relationships that could have appeared to influence the work reported in this paper.

\section*{Funding}
This work was supported by the Khalifa University of Science and Technology under Award RIG-2023-117, HARBOT: KU\_UB JRP, and KUCARS.

\makeatletter
\setlength{\bibsep}{0pt plus 0.3ex}  
\makeatother

\end{document}